\documentclass[a4paper,11pt]{article}

\pdfoutput=1
\usepackage{jheppub}

\usepackage{multirow}
\usepackage{titlesec}
\usepackage{float}
\titleformat*{\subsection}{\bfseries\boldmath}

\usepackage[T1]{fontenc}

\usepackage[caption=false]{subfig}

\newcommand{\pphi}{\beta}
\newcommand{\gphi}{\gamma}

\newcommand{\ckm} {\mathrm{CKM}}
\newcommand{\km} {\mathrm{KM}}
\newcommand{\ut} {\mathrm{UT}}

\newcommand{\lbr}{\left(}
\newcommand{\rbr}{\right)}

\newcommand{\dt}{\Delta t}
\newcommand{\mpsq}{m_+^2}
\newcommand{\mmsq}{m_-^2}

\newcommand{\dmb}{\Delta m_B}
\newcommand{\dmdt}{\left(\dmb\dt\right)}
\newcommand{\hdmdt}{\lbr\frac{\dmb\dt}{2}\rbr}

\newcommand{\dkap}{\varkappa\lbr t_D \rbr}
\newcommand{\dsig}{\sigma\lbr t_D \rbr}

\newcommand{\dn}{D^0}
\newcommand{\bn}{B^0}
\newcommand{\dbar}{\overline{D}}
\newcommand{\bbar}{\overline{B}}
\newcommand{\dnbar}{\dbar{}^0}
\newcommand{\bnbar}{\bbar{}^0}
\newcommand{\ks}{K_S^0}
\newcommand{\pipi}{\pi^+\pi^-}
\newcommand{\bdpp}{\bn\to D\pipi}
\newcommand{\bdnpp}{\bn\to \dn\pipi}
\newcommand{\bbdnpp}{\bnbar\to \dn\pipi}
\newcommand{\kspp}{\ks\pipi}
\newcommand{\dkpp}{D\to\kspp}
\newcommand{\dnkpp}{\dn   \to \kspp}
\newcommand{\dbkpp}{\dnbar\to \kspp}
\newcommand{\bdh}{\bn\to D h^0}

\newcommand{\btodh}{\bn\to D h^0}
\newcommand{\bdsth}{\bn\to D{}^{(*)}h^0}

\newcommand{\ifb}{~\mathrm{fb}^{-1}}

\newcommand{\cpconj}{\mathcal{CP}}
\newcommand{\cconj}{\mathcal{C}}

\newcommand{\bdcph}{\bn\to D_{\cpconj} h^0}
\newcommand{\bdkspph}{\bn\to [\kspp]_{D} h^0}

\newcommand{\bdcppp}{\bn\to D_{\cpconj}\pipi}
\newcommand{\bdkspppp}{\bn\to [\kspp]_{D}\pipi}

\newcommand{\etal}{\textit{et al.}}

\newcommand{\bdk}{B^{\pm}\to DK^{\pm}}

\newcommand{\ki}{K_i}
\newcommand{\kibar}{\overline{K}_i}
\newcommand{\kmi}{K_{-i}}

\newcommand{\ci}{C_i}
\newcommand{\cmi}{C_{-i}}
\newcommand{\si}{S_i}
\newcommand{\smi}{S_{-i}}

\newcommand{\kj}{k_j}
\newcommand{\kmj}{k_{-j}}

\newcommand{\cj}{c_j}
\newcommand{\cmj}{c_{-j}}
\newcommand{\sj}{s_j}
\newcommand{\smj}{s_{-j}}
\newcommand{\djth}{d_j}

\newcommand{\kb}{\overline{k}}
\newcommand{\kbj}{\kb_j}
\newcommand{\kbmj}{\kb_{-j}}

\newcommand{\cbar}{\overline{c}}
\newcommand{\sbar}{\overline{s}}
\newcommand{\cbj}{\cbar_j}
\newcommand{\cbmj}{\cbar_{-j}}
\newcommand{\sbj}{\sbar_j}
\newcommand{\sbmj}{\sbar_{-j}}

\newcommand{\ct}{\widetilde{c}}
\newcommand{\st}{\widetilde{s}}
\newcommand{\ctj}{\ct_j}
\newcommand{\ctmj}{\ct_{-j}}
\newcommand{\stj}{\st_j}
\newcommand{\stmj}{\st_{-j}}

\newcommand{\cprm}{c^{\prime}}
\newcommand{\sprm}{s^{\prime}}
\newcommand{\cpj}{\cprm_j}
\newcommand{\cpmj}{\cprm_{-j}}
\newcommand{\spj}{\sprm_j}
\newcommand{\spmj}{\sprm_{-j}}

\newcommand{\mev}{\mathrm{MeV}}
\newcommand{\gev}{\mathrm{GeV}}
\newcommand{\mevc}{\mev/c}
\newcommand{\gevc}{\gev/c}
\newcommand{\mevcsq}{\mevc^{2}}
\newcommand{\gevcsq}{\gevc^{2}}

\newcommand{\btoccs}{b\to c\overline{c}s}

\newcommand{\btocud}{b\to c\overline{u}d}
\newcommand{\btoucd}{b\to u\overline{c}d}

\newcommand{\dbpp}{\dnbar\pipi}
\newcommand{\dpp}{\dn\pipi}

\newcommand{\abdbpp}{\mathcal{A}_{\dbpp}}
\newcommand{\abbdpp}{\overline{\mathcal{A}}{}_{\dpp}}

\newcommand{\sindbeta}{\sin{2\pphi}}
\newcommand{\cosdbeta}{\cos{2\pphi}}
\newcommand{\grad}    {{}^{\circ}}

\newcommand{\belle}{\mathrm{Belle}}
\newcommand{\belleii}{\belle\,\mathrm{II}}
\newcommand{\bellebii}{\belle\,(\mathrm{II})}

\newcommand{\lhcb}{\mathrm{LHCb}}

\newcommand{\etag}{\varepsilon_{\mathrm{tag}}}

\newcommand{\mcf}{\mathcal{F}}

\newcommand{\mcn}{\mathcal{N}}
\newcommand{\mcm}{\mathcal{M}}
\newcommand{\mcr}{\mathcal{R}}

\newcommand{\mca}{\mathcal{A}}
\newcommand{\mcb}{\mathcal{B}}

\newcommand{\mcab}{\overline{\mathcal{A}}}
\newcommand{\ab}{\mca_B}
\newcommand{\ad}{\mca_D}
\newcommand{\af}{\mca_f}
\newcommand{\abbar}{\mcab_B}
\newcommand{\adbar}{\mcab_D}

\newcommand{\bdbpp}{\bn\to\dnbar\pipi}

\newcommand{\abf}{\overline{\mathcal{A}}{}_f}
\newcommand{\afb}{\mathcal{A}_{\overline{f}}}
\newcommand{\abfb}{\overline{\mathcal{A}}{}_{\overline{f}}}

\newcommand{\dvar}   {\left(\mpsq,\mmsq\right)}
\newcommand{\dvarinv}{\left(\mmsq,\mpsq\right)}
\newcommand{\bvar}   {\left(\mu^2_{+},\mu^2_{-}\right)}
\newcommand{\bvarinv}{\left(\mu^2_{-},\mu^2_{+}\right)}

\newcommand{\ddvar}   {\,\mathrm{d}\mpsq\mathrm{d}\mmsq}

\newcommand{\dtbvardvar} {\lbr\dt,\mu^2_{+},\mu^2_{-},\mpsq,\mmsq\rbr}

\newcommand{\ucoef}{\mathcal{U}}
\newcommand{\ccoef}{\mathcal{D}}
\newcommand{\scoef}{\mathcal{F}}
\newcommand{\ccoefcp}{\ccoef^{\cpconj}_j}
\newcommand{\scoefcp}{\scoef^{\cpconj}_j}
\newcommand{\ccoefdd}{\ccoef_{ij}}
\newcommand{\scoefdd}{\scoef_{ij}}
\newcommand{\ccoefcpsym}{\ccoef^{\cpconj}_{|j|}}
\newcommand{\scoefcpsym}{\scoef^{\cpconj}_{|j|}}
\newcommand{\ccoefddsym}{\ccoef_{i|j|}}
\newcommand{\scoefddsym}{\scoef_{i|j|}}

\newcommand{\ddel}{\Delta\delta}
\newcommand{\deld}{\ddel_D}
\newcommand{\delb}{\ddel_B}

\newcommand{\cfcoef}{\ccoef_f}
\newcommand{\sfcoef}{\scoef_f}

\newcommand{\cflvcoef}{\ccoef_{\mathrm{flv}}}
\newcommand{\sflvcoef}{\scoef_{\mathrm{flv}}}

\newcommand{\ccpcoef}{\ccoef_{\cpconj}}
\newcommand{\scpcoef}{\scoef_{\cpconj}}

\newcommand{\ccoefbdpp}{\ccoef_{\dpp}}
\newcommand{\scoefbdpp}{\scoef_{\dpp}}

\newcommand{\btau}{\tau_B}
\newcommand{\bexp}{e^{-\frac{\left|\Delta t\right|}{\btau}}}
\newcommand{\lamfmsq}{\left|\lambda_f\right|^2}
\newcommand{\lamf}{\lambda_f}

\newcommand{\ep}{e^+e^-}
\newcommand{\pin}{\pi^0}

\newcommand{\dm}{\Delta m_B}

\newcommand{\ith}{{^{\mathrm{th}}}}

\newcommand{\fcp}{f_{\cpconj}}

\newcommand{\pbar}{\overline{p}}

\newcommand{\runi}{\mathrm{Run\ I}}
\newcommand{\runii}{\mathrm{Run\ II}}
\newcommand{\upg}{\mathrm{Upgr.}}

\newcommand{\imag}{\mathrm{Im}}
\newcommand{\real}{\mathrm{Re}}

\newcommand{\subbkg}{_{{\rm bkg}}}

\newcommand{\abd}{\mca_{B\to D}}
\newcommand{\abdbar}{\mca_{B\to\dbar}}

\newcommand{\pbd}{p_{B\to D}}
\newcommand{\pbdbar}{p_{B\to\dbar}}

\newcommand{\abbard}{\mca_{\bbar\to D}}
\newcommand{\abbardbar}{\mca_{\bbar\to\dbar}}
\newcommand{\rbd}{r_{B}}
\newcommand{\dbd}{\Delta\delta_{B}}

\newcommand{\psib}{\psi_{B}}
\newcommand{\psibb}{\psi_{\bbar}}

\title{\boldmath A method for model-independent measurement of the $\ckm$ 
angle $\pphi$ via time-dependent analysis of the $\bdpp$, $\dkpp$ decays}

\author[a,b]{A.~Bondar,}
\author[a,b]{A.~Kuzmin,}
\author[a,b,c]{V.~Vorobyev}

\affiliation[a]{Novosibirsk State University,\\
                Pirogova st. 2, 630090, Novosibirsk, Russia}
\affiliation[b]{Budker Institute of Nuclear Physics SB RAS,\\
                Lavrentiev ave. 11, 630090, Novosibirsk, Russia}
\affiliation[c]{P.N. Lebedev Physical Institute of the Russian Academy of Sciences,\\
                Leninskii pr. 53, 119991, Moscow, Russia}

\emailAdd{vvorob@inp.nsk.su}

\abstract{
 A new method for model-independent measurement of the~$\ckm$ 
 angle~$\pphi$ is proposed, that employs time-dependent analysis of 
 flavour-tagged~$\bdpp$ decays with~$D$ meson decays into 
 $\cpconj$-specific and~$\kspp$ final states.  This method can be used 
 to measure the angle~$\pphi$ with future data from the~$\belleii$ 
 and~$\lhcb$ experiments with the precision level of one degree.
}

\begin{document} 
\maketitle
\flushbottom

\section{Introduction}\label{sec:intro}
The $B$-factory experiments at SLAC~\cite{babar} and KEK~\cite{belle} have 
made impressive progress in studies of the $\cpconj$ symmetry breaking 
in~$B$ meson decays. The~$\lhcb$~\cite{lhcb} experiment has been contributing 
significantly to this field since recently.  The $\cpconj$-violating 
phenomena observed so far are in agreement with the $\km$~mechanism of 
the~$\cpconj$ symmetry breaking proposed by Cabibbo, Kobayashi and
Maskawa~\cite{cabibbo,KM}.  Nevertheless, theoretical estimates~\cite{baryogenesis} 
claim that the $\km$ mechanism cannot provide the value of $\cpconj$ violation 
large enough to generate the observed baryon asymmetry of the Universe~\cite{sakharov}.  
Thus, searches for other mechanisms of $\cpconj$ violation and tests of the~$\km$ 
mechanism should be continued.  

Comparison of the angle~$\pphi$ values of the Unitarity Triangle~($\ut$)~\cite{ut}
measured in different processes is a valuable test of the $\km$~mechanism.
The value of~$\sindbeta$ obtained using the~$\btoccs$ 
transitions~\cite{sinbeta_first1, sinbeta_first2, sinbeta_babar, sinbeta_belle, 
sinbeta_lhcb} is currently the most precisely measured parameter related to the UT angles~\cite{hfag}:
\begin{equation}\label{eq:sindbeta_ccs}
 \sindbeta^{(\btoccs)} = 0.691 \pm 0.017.
\end{equation}
The value of~$\sindbeta$ measured in the~$\btocud$ transitions~\cite{sinbeta_bcud} is consistent with the~$\btoccs$ result though it is statistically limited:
\begin{equation}\label{eq:sindbeta_cud}
 \sindbeta^{(\btocud)} = 0.66 \pm 0.10 \pm 0.06.
\end{equation}

\begin{figure}
 \centering
 \subfloat[]{\label{fig:bccs}%
  \includegraphics[width=0.22\textwidth]{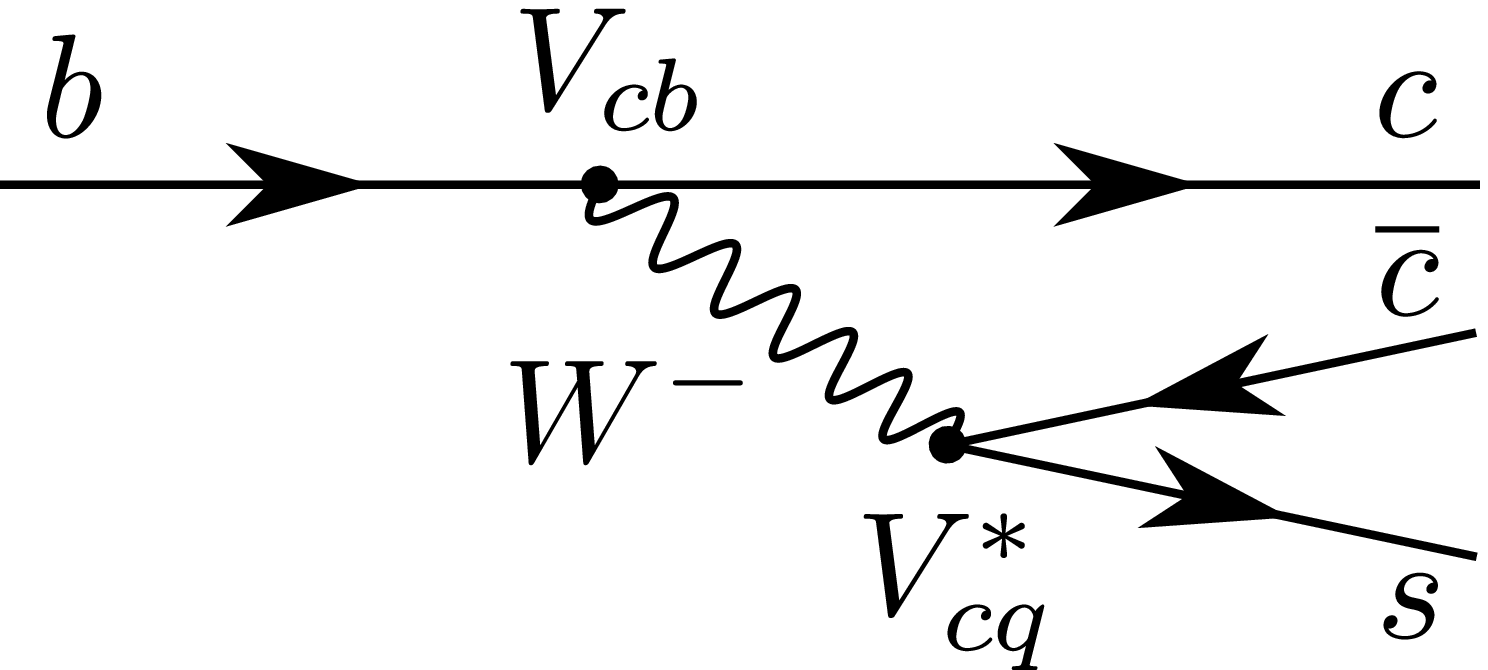}} \hspace{0.03\textwidth}
 \subfloat[]{\label{fig:bccs_loop}%
  \includegraphics[width=0.22\textwidth]{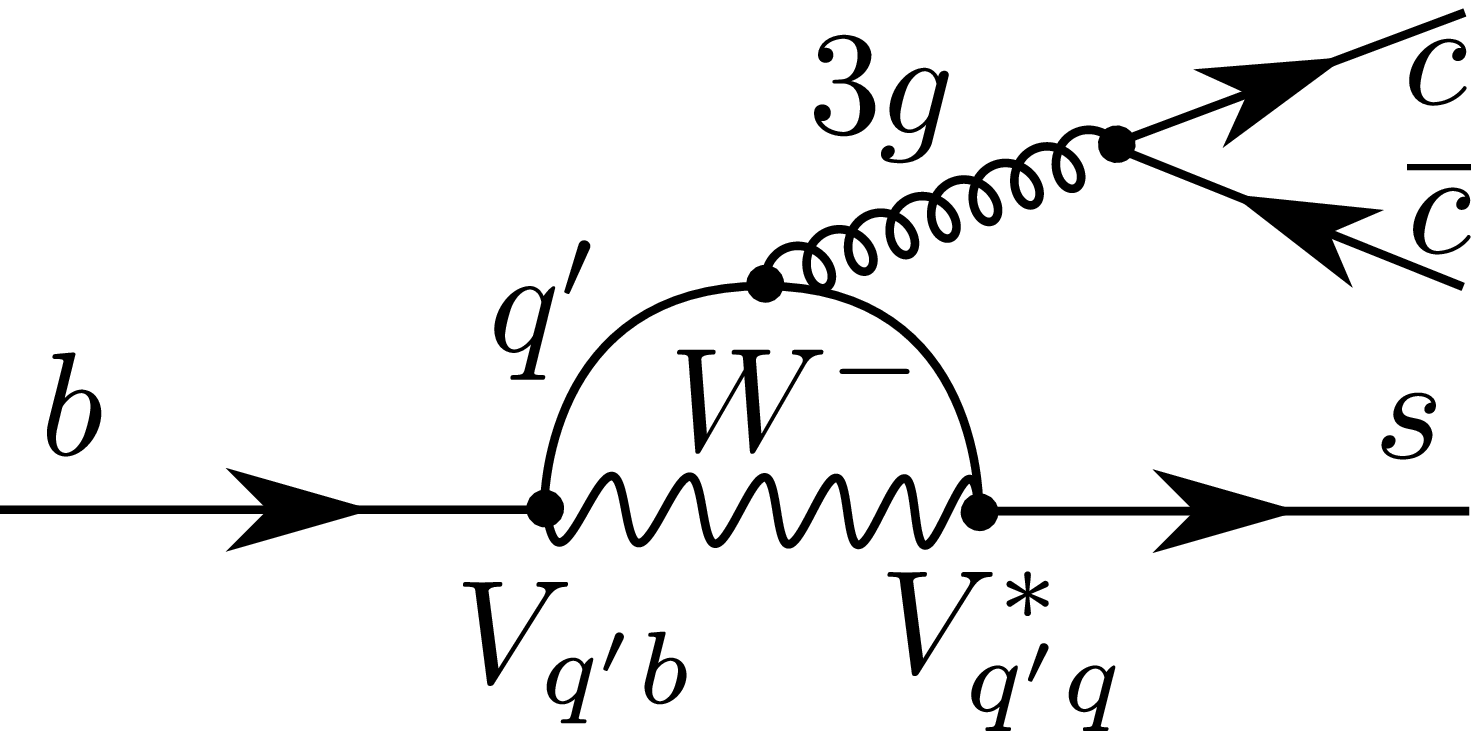}}  \hspace{0.03\textwidth}
 \subfloat[]{\label{fig:bcud}%
  \includegraphics[width=0.22\textwidth]{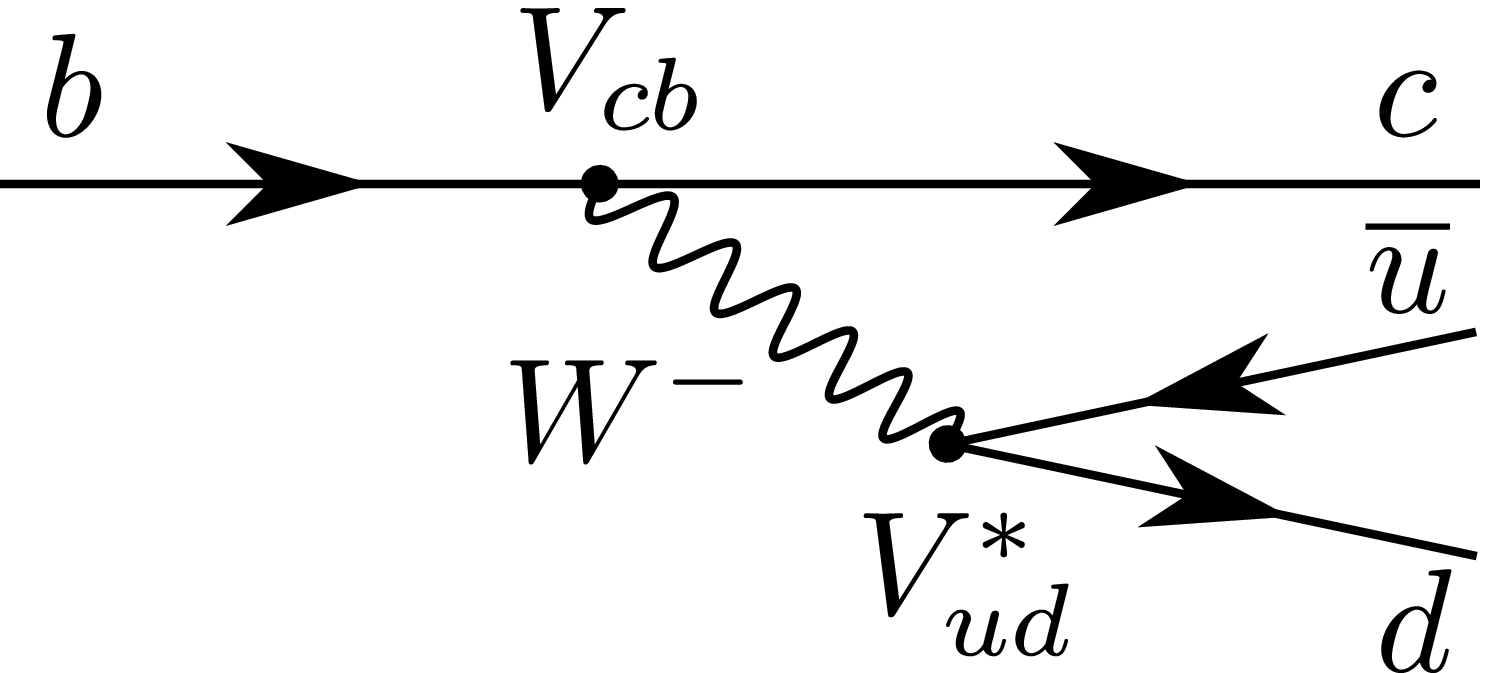}} \hspace{0.03\textwidth}
 \subfloat[]{\label{fig:bucd}%
  \includegraphics[width=0.22\textwidth]{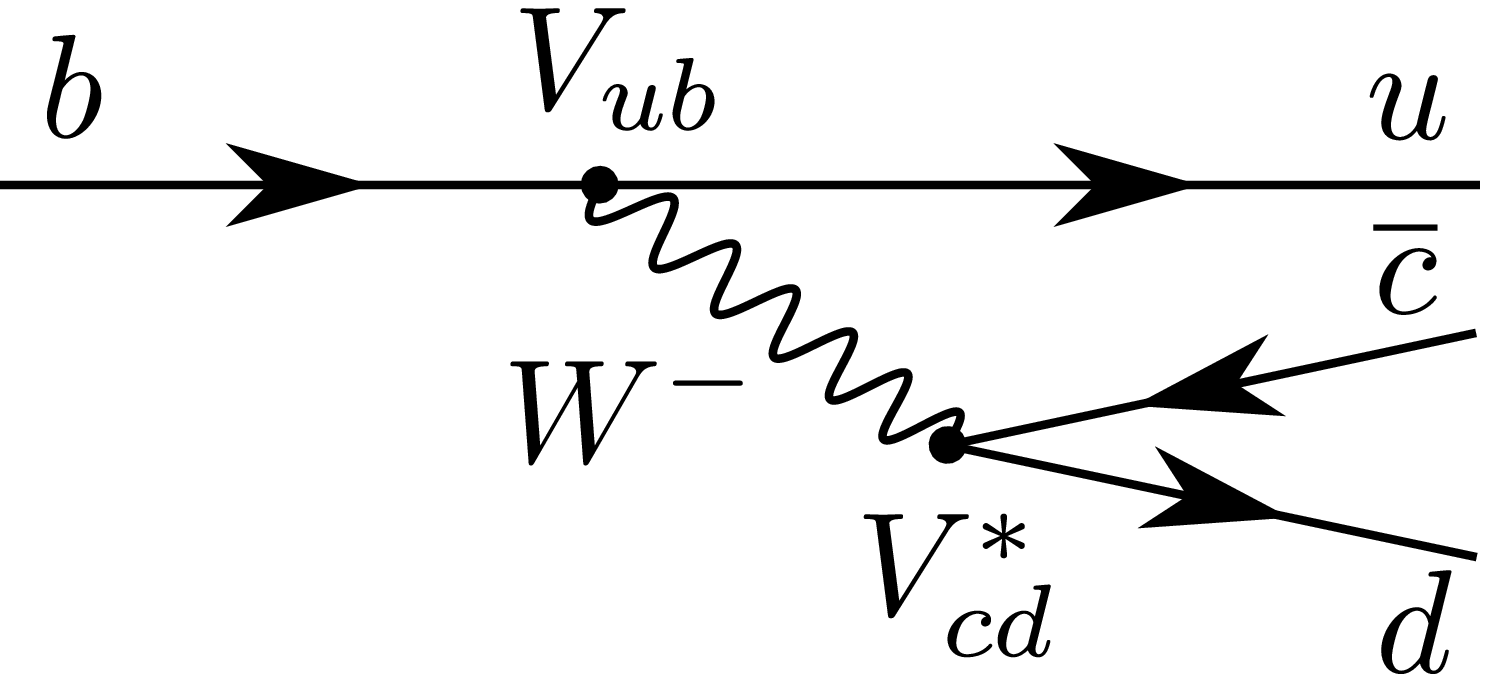}}
 \caption{The tree~(a) and loop~(b) contributions to the $\btoccs$ transition, 
 $\btocud$ transition~(c) and suppressed $\btoucd$ transition~(d).}
 \label{fig:btransitions}
\end{figure}

Within the Standard Model, the angle $\pphi$ measurements in~$\btoccs$ and~$\btocud$ 
transitions should give the same result up to the hadronic corrections that are expected to be small. However, due to the difference 
of the~$\btoccs$ and~$\btocud$ structure (see Figure~\ref{fig:btransitions}), 
the New Physics phenomena may manifest themselves differently in these 
transitions~\cite{acp_np}. The doubly Cabibbo-suppressed loop contributions 
to the $\btoccs$ transitions, limiting the interpretation of measurements, 
can be controlled using the~$SU(3)$ flavor symmetry, as it is shown by De Bruyn 
and Fleischer in Ref.~\cite{bruyn_fleischer}. Bias of the observable~$2\pphi$ 
value can be controlled at the level of~$0.3\grad$ assuming~$20\%$ accuracy in  
$U$-symmetry approximation.

The obtained value of~$\sindbeta$ leaves the ambiguity $\pphi\to\pi/2 - \pphi$, 
which can be resolved by measuring $\cosdbeta$.  Several approaches to 
measure $\cosdbeta$ in the $\btocud$ transitions using the time-dependent Dalitz 
plot analysis were discussed: (1) the analysis of $\bdh$, $\dkpp$ decays was proposed 
in Ref.~\cite{gershon_bondar_krokovny}, (2) the analysis of $\bn\to D_{\cpconj}\pipi$ 
decays was mentioned in Ref.~\cite{cyopr} and considered in detail in 
Ref.~\cite{latham_gershon}, (3) the analysis of $\bdpp$, $\dkpp$ decays was mentioned 
in Ref.~\cite{latham_gershon}. Only the $\bdh$, $\dkpp$ decays analysis was 
implemented in practice providing the first~\cite{cosbeta_first_result} 
as well as the most precise at the moment measurements of~$\cosdbeta$~\cite{cosbeta_babar, cosbeta_belle}.\footnote{Results of the $\cosdbeta$ measurement in $\bdh$, $\dkpp$ decays via joint analysis of the Belle and BaBar experiments data are being prepared for publication at the moment. It is expected to be the most precise measurement of $\cosdbeta$ before the Belle II data is available. See the talk by M.~Roehrken at the 52nd Rencontres de Moriond EW 2017 conference.}
These results indicate positiveness of the~$\cosdbeta$ as expected within 
the~$\km$ mechanism.

Measurements of~$\cosdbeta$ in~$\bdh$, $\dkpp$ decays 
require knowledge of the phase difference~$\deld$ between the amplitudes 
of~$\dbkpp$ and~$\dnkpp$ decays that varies over the phase space and cannot 
be measured directly.  The common workaround is to build a phenomenological 
decay amplitude model and obtain the $D$ meson decay amplitude phase from the 
model.  A model uncertainty is inherent in this approach.

The~$\lhcb$ and~$\belleii$~\cite{belleII} experiments are expected to collect 
samples of~$B$ meson decays much larger than those available today.  Precision 
of model-dependent measurements of the angle~$\pphi$ with that statistics will 
probably be limited by the model uncertainty.  Indeed, currently the model 
uncertainty is assessed mostly from the statistical error of model parameters, 
assuming that the obtained value exceeds the uncertainty related to justification 
of the model approach.  There is no reason to rely on this assumption in a 
percent-precision-level measurement.

The idea of binned Dalitz plot analysis proposed in~Ref.~\cite{GGSZ} was 
to overcome the limitations of model-dependent consideration of 
multibody decays. The initial idea is related to measuring the~$\ut$ angle 
$\gphi$ in $\bdk$, $\dkpp$ decays.  It was developed further and extended 
to several other applications in~Refs.~\cite{BP_phi3_model1,BP_phi3_model2,BPV,dd,hr1,hr2,BDKpiModInd,UnbinnedModInd}.  
A measurement of~$\cosdbeta$ in~Ref.~\cite{cosbeta_belle} has been performed 
in a model-independent way using these ideas.

In this work, the model-independent approach is considered in a context of
the angle~$\pphi$ measurement in time-dependent analysis of~$\bdpp$ decays 
with~$D$ meson decaying into~$\cpconj$-specific and~$\kspp$ states.  
It is shown the angle~$\pphi$ and necessary hadronic parameters 
of the $\bdbpp$ decay can be obtained in a single measurement.  Formalism of the 
time-dependent analysis of the $\bdpp$ decays is described in Sec.~\ref{sec:pdfs}.  
The method for model-independent measurement of the angle~$\pphi$ with the $\bdpp$ 
decays is developed in~Sec.~\ref{sec:bins}.  The statistical precision 
with future data of the $\belleii$ and $\lhcb$ experiments 
is evaluated in~Sec.~\ref{sec:feas}.  The measurement bias due to the neglect of 
$\btocud$ transition and charm mixing is considered in appendices~\ref{app:bucbard}, 
\ref{app:charm-mixing}, and \ref{app:systematics-evaluation}.

\section{\boldmath Time-dependent analysis of \texorpdfstring{$\bdpp$}{B0 to 
anti-D0 pi+ pi-} decays}\label{sec:pdfs}
Phenomenology of time-dependent $\cpconj$ violation measurements at an 
asymmetric-energy $\ep$ $B$-factory is described elsewhere~\cite{BPhys}.  
The decay probability density for a flavour-tagged $B$ meson is expressed by
\begin{equation}\label{eq:raw_pdf}
  p(\dt) \propto \bexp\left[1 +q_B\lbr\cfcoef\cos{\dmdt}-\sfcoef\sin{\dmdt}\rbr\right],
\end{equation}
where $\dt\in\lbr-\infty, \infty\rbr$ is the proper decay time of a tagged $B$ 
meson counted from the moment of the tagging $B$ meson decay,\footnote{Corresponding 
expressions for the time-dependent analysis at~$\lhcb$ are obtained by the formal 
substitution of~$\dt\to t$, where~$t\in[0, \infty)$.} 
$q_B = 1$~($q_B = -1$) corresponds to~$\bn$~($\bnbar$) flavour at $\dt = 0$, 
$\dm$ is the mass difference between the $B$ meson mass eigenstates, $\btau$
is the $\bn$ lifetime, and
\begin{equation}
  \cfcoef = \frac{1-\lamfmsq}{1+\lamfmsq},\quad 
  \sfcoef = \frac{2\,\imag\,\lamf}{1+\lamfmsq},
\end{equation}
where $f$ denotes the $B$ meson final state and
\begin{equation}
  \lamf=\frac{q}{p}\frac{\abf}{\af},
\end{equation}
where $q$ and $p$ are the parameters of $B$ meson mixing and $\af$ ($\abf$) 
is the $B^0\to f$ ($\bnbar\to f$) decay amplitude.  Hereafter, absence of 
direct $\cpconj$ symmetry breaking in $B$ and $D$ meson decays as well as 
absence of $\cpconj$ symmetry breaking in $B$ meson mixing are assumed\footnote{
The case of direct $\cpconj$ violation in $B$ meson decay due to
the $\btoucd$ quark transition is considered in Appendix~\ref{app:bucbard}. 
The effect of charm mixing is considered in Appendix~\ref{app:charm-mixing}.}
which implies
\begin{equation}
  \frac{q}{p} = e^{-2i\pphi},\quad \af\equiv\abfb,\quad \abf\equiv\afb,
\end{equation}
where $\overline{f}$ denotes the state obtained by $\cpconj$ 
conjugation of state~$f$.

The amplitude of $\bdbpp$, $\dnbar\to f_D$ can be expressed as
\begin{equation}
 \abdbpp \propto \ab\bvar\adbar,
\end{equation}
where $\adbar$ is the $\dnbar$ meson decay amplitude and $\ab$ depends on the 
Dalitz plot variables~$\mu_{\pm}^2 \equiv m^2\lbr D\pi^{\pm}\rbr$. 

The amplitude of the $\cpconj$-conjugated process, $\bnbar\to\dn\pipi$, 
$\dn\to\overline{f}_D$, is
\begin{equation}
 \abbdpp \propto \abbar\bvar\ad\equiv\ab\bvarinv\ad.
\end{equation}
The parameters $\cfcoef$ and $\sfcoef$ from Eq.~(\ref{eq:raw_pdf}) 
take the form
\begin{subequations}\label{eq:y}
 \begin{align}
  \label{eq:y:1}
  \ccoefbdpp &=
    \frac{p_B\bvar\pbar_D - p_B\bvarinv p_D}
         {p_B\bvar\pbar_D + p_B\bvarinv p_D}, \\ 
  \label{eq:y:2}
  \scoefbdpp &=
    \frac{2\sqrt{p_B\bvar\pbar_Dp_B\bvarinv p_D}}
         {p_B\bvar\pbar_D + p_B\bvarinv p_D}\times\sin{\lbr\ddel_f-2\pphi\rbr},
 \end{align}
\end{subequations}
where $p_B = \left|\ab\right|^2$, $p_D = \left|\ad\right|^2$, 
$\pbar_D = \left|\adbar\right|^2$, $\ddel_f = \ddel_B - \ddel_D$ and
\begin{equation}\label{eq:phases-definition}
 \delb\bvar = \arg{\left(\frac{\ab\bvarinv}{\ab\bvar}\right)},\ 
 \deld = \arg{\left(\frac{\adbar}{\ad}\right)}.
\end{equation}

If the $D$ meson is reconstructed in a flavour-specific final state, then 
$\cflvcoef = 1$ and $\sflvcoef = 0$.\footnote{Hadronic decays like 
$\dn\to K^-\pi^+$ are used in practice instead of flavour-specific decays.  
The relations $\ccoef = 1$ and $\scoef = 0$ do not hold in this case 
because of suppressed decays $\dnbar\to K^-\pi^+$.  The suppressed decays 
can be taken into account in a high-statistics measurement.} 
A $\cpconj$-specific $D$ meson final 
state with $\cpconj$ parity $\xi_D$ results in
\begin{subequations}\label{eq:raw_pdf_cp}
 \begin{align}
  \ccpcoef & =  \frac{p_B\bvar - p_B\bvarinv}{p_B\bvar + p_B\bvarinv}, \\ 
  \scpcoef & =  \frac{2\sqrt{p_B\bvar p_B\bvarinv}}
                          {p_B\bvar + p_B\bvarinv}\times \xi_D \sin{(\ddel_B-2\pphi)}.
 \end{align}
\end{subequations}
The final state $\kspp$ introduces the second Dalitz plot resulting in 
dependence of the $D$ meson decay probability density and the phase difference 
between the $\dnbar$ and $\dn$ decay amplitudes on the Dalitz plot variables 
$m_{\pm}^2 = m^2\lbr\ks\pi^{\pm}\rbr$:
\begin{equation}\label{eq:dcpcons}
 \pbar_D\dvar \equiv p_D\dvarinv,\quad \ddel_D\dvar.
\end{equation}
In this case, the $B$ meson decay probability density from Eq.~(\ref{eq:raw_pdf}) 
depends on time and four Dalitz plot variables.

In principle, any multibody self-conjugated final state, such as $\ks K^+K^-$, 
$\pipi\pin$ or $K^+K^-\pipi$ can be considered, but the $\kspp$ state is 
the most experimentally clean and has rich resonance structure leading to 
significant variation of the phase difference~$\deld$ over the Dalitz plot and
good sensitivity to the $\cpconj$ violation parameters.  
Similar formalism can be developed for other multibody hadronic $D$~meson 
final states, such as $K^-\pi^+\pin$.  The $D$~meson decay probability 
densities $p_D$ and $\pbar_D$ would be independent in that case.

\section{Binned Dalitz plot analysis}\label{sec:bins}
The decay probability densities derived in the previous section can be 
expressed in terms of the parameters of the binned Dalitz plot. We 
follow the notation introduced in~Ref.~\cite{BPV}, where the~$\dnkpp$ 
Dalitz plot is divided into~$2\mcn$ bins (we use $\mcn=8$).  The partitioning 
is done so that the bin index $i$ ranges from $-\mcn$ to $\mcn$ excluding 
zero and the sign inversion $i\to -i$ corresponds to the Dalitz plot 
reflection~$\mpsq\leftrightarrow\mmsq$.  The parameters~$\ki$, $\kibar$, 
$\ci$ and~$\si$ are defined for the~$i\ith$ bin:
\begin{equation}
 \ki    \equiv \frac{\int\limits_{\mathcal{D}_i}p_D\ddvar}
                    {\sum\limits_i{\int\limits_{\mathcal{D}_i}p_D\ddvar}},\quad
 \kibar \equiv \frac{\int\limits_{\mathcal{D}_i}\pbar_D\ddvar}
                    {\sum\limits_i \int\limits_{\mathcal{D}_i}\pbar_D\ddvar},\quad
 \ci    \equiv \real\,e_i,\quad \si \equiv \imag\,e_i,
\end{equation}
where integration is performed over the $i\ith$ bin and
\begin{equation}
 e_i \equiv \frac{\int\limits_{\mathcal{D}_i}\ad^{*}\dvar\ad\dvarinv\ddvar}
            {\sqrt{\int\limits_{\mathcal{D}_i}p_D\dvar\ddvar}
           \ \sqrt{\int\limits_{\mathcal{D}_i}p_D\dvarinv\ddvar}}.
\end{equation}
The relation~(\ref{eq:dcpcons}) and symmetry of the Dalitz plot 
partitioning lead to the relations $\ci \equiv \cmi$, $\si \equiv -\smi$, and 
$\kibar \equiv \kmi$.

In a similar way, we divide the~$\bdpp$ decay Dalitz plot into 
$2\mcm=2\times8$ bins and define the parameters $\kj$, $\cj$ and $\sj$ 
for that Dalitz plot, where the bin index~$j$ ranges from $-\mcm$ to 
$\mcm$ excluding zero.  A time-dependent $\bdpp$ decay probability density
\begin{equation}\label{eq:binned_pdf}
 N_{j}(\dt) \propto \bexp  \left[1 + q_B\ccoef_{j}\cos{\dmdt}-
                                     q_B\scoef_{j}\sin{\dmdt}\right],
\end{equation}
is defined for the~$j^{\textrm{th}}$ bin.  In the case of double Dalitz decay
$\bdpp$, $\dkpp$, the decay probability density is defined for each combination 
of $\bn$ Dalitz plot bin $j$ and $\dn$ Dalitz plot bin $i$:
\begin{equation}\label{eq:double_binned_pdf}
 N_{ij}(\dt) \propto \bexp \left[1 + q_B\ccoef_{ij}\cos{\dmdt}-
                                     q_B\scoef_{ij}\sin{\dmdt}\right].
\end{equation}

The following substitutions are used to express the coefficients~$\ccoef$ 
and~$\scoef$ in the form suitable for the binned analysis:
\begin{subequations}
 \begin{align}
  p_B\bvar \to \kj,&\quad p_B\bvarinv \to \kmj,\\
  p_D\dvar \to \ki,&\quad p_D\dvarinv \to \kmi,\\
  \sin{\deld} \to \si,&\quad \cos{\deld} \to \ci,\\
  \sin{\delb} \to \sj,&\quad \cos{\delb} \to \cj.
 \end{align}
\end{subequations}

The expression Eq.~(\ref{eq:raw_pdf_cp}) for the $\cpconj$-specific $D$ meson 
decays transforms into
\begin{subequations}\label{eq:cp_pdf}
\begin{align}
 \ccoefcp & = \frac{\kj-\kmj}{\kj+\kmj},\\
 \scoefcp & = 2\xi_D\frac{\sqrt{\kj\kmj}}{\kj+\kmj}\lbr\sj\cosdbeta-
                                                       \cj\sindbeta\rbr.
\end{align}
\end{subequations}
The double Dalitz plot case with the $\dnkpp$ decay results in
\begin{subequations}\label{eq:dd_pdf}
\begin{align}
 \ccoefdd & = \frac{\ki\kj-\kmi\kmj}{\ki\kj+\kmi\kmj},\\
 \scoefdd & = 2\frac{\sqrt{\ki\kmi\kj\kmj}}{\ki\kj+\kmi\kmj}
           \times \left[\lbr\ci\sj-\si\cj\rbr\cosdbeta-
                        \lbr\ci\cj+\si\sj\rbr\sindbeta\right].
\end{align}
\end{subequations}

We consider the parameters $\ki$, $\ci$ and $\si$ to be known because they 
can be measured in decays of coherent~$\dn\dnbar$ pairs~\cite{CLEO_phasees}.  
The $2\mcm$~parameters~$\kj$, $\mcm$ parameters~$\cj$, $\mcm$~parameters~$\sj$ 
and the angle~$\pphi$ constitute $4\mcm+1$ unknown parameters.  

The parameters $\kj$ can be measured precisely in the time-integrated analysis 
of $\bbdnpp$ decays with $\dn$~meson decaying into hadronic state~$K^-\pi^+$.  
The expected fraction of events in the $j^{\mathrm{th}}$ Dalitz plot bin is
\begin{equation}\label{eq:kj}
 N_j \approx \kj - r_D^2\frac{1-z}{1+z}\lbr\kj - \kmj\rbr,
\end{equation}
where
\begin{equation}
 z\equiv \frac{1}{1 + \lbr\dm\btau\rbr^2}\approx 0.6,\quad
 r_D^2\equiv \frac{Br\lbr\dn\to K^+\pi^-\rbr}
                  {Br\lbr\dn\to K^-\pi^+\rbr}\approx 3.5\times 10^{-3}.
\end{equation}
The second term in Eq.~(\ref{eq:kj}) is negligible even at the $\belleii$ 
precision level.

The $\bdpp$ with $\cpconj$-specific $D$~meson decays provide~$2\mcm$ 
independent constraints (Eq.~(\ref{eq:cp_pdf})) and do not allow one to 
resolve the system.  It should be noted that the above statement does 
not depend on $\cpconj$ parity of the $D$ meson final state, particularly, 
final states with the same $\cpconj$ parities can be used and inclusion of 
a final state of the opposite $\cpconj$ parity would not increase the 
number of constraints.

The $\bdpp$ with $\dkpp$ decay provide $2\mcm\mcn$ additional constraints 
(Eq.~(\ref{eq:dd_pdf})) allowing to measure the parameters $\cj$ and 
$\sj$ together with the angle $\pphi$ in the joint analysis of the 
$\bdpp$ with $\cpconj$-specific and $\dkpp$ decays for 
any $\mcn$ and $\mcm$.\footnote{
An important feature of the described setup is that the values of 
$\sindbeta$ and $\cosdbeta$ cannot be considered as independent parameters. 
Indeed, the transformation
\begin{equation}\label{eq:scale}
  \cj\to \eta\cj,\quad \sj\to \eta\sj,\quad
  \sindbeta\to\frac{\sindbeta}{\eta},\quad\cosdbeta\to\frac{\cosdbeta}{\eta}
\end{equation}
with an arbitrary scale~$\eta\neq 0$ does not change the expressions for 
decay probability densities and the scale~$\eta$ can not be determined.}
The $\bdpp$, $\dkpp$ decays alone provide enough constraints 
to measure the parameters~$\cj$ and~$\sj$, and the angle~$\pphi$ 
for~$2\mcm(\mcn-1)\geq1$.

\subsection{Symmetrized \texorpdfstring{$\bdbpp$}{B0 -> anti-D0 pi+ pi-} 
Dalitz plot binning}\label{sec:symm_dp}
The number of parameters related to the $\bdbpp$ binned Dalitz plot can be 
reduced by a factor of~$2$ considering the $j^{\mathrm{th}}$ and 
$-j^{\mathrm{th}}$ bins as a single bin.  For the symmetrized in this way 
$\bdbpp$ decay Dalitz plot binning, the expressions Eq.~(\ref{eq:cp_pdf}) and 
Eq.~(\ref{eq:dd_pdf}) should be modified as follows:
\begin{equation}\label{eq:cp_dil_pdf}
 \ccoefcpsym = 0,\quad \scoefcpsym = \djth\sindbeta
\end{equation}
and
\begin{equation}\label{eq:dd_dil_pdf}
 \ccoefddsym = \frac{\ki-\kmi}{\ki+\kmi},\quad
 \scoefddsym =-2\djth\frac{\sqrt{\ki\kmi}}{\ki+\kmi}
               \lbr\si\cosdbeta+\ci\sindbeta\rbr,
\end{equation}
where the \emph{dilution factor}
\begin{equation}\label{eq:dilut}
 \djth = 2\frac{\sqrt{\kj\kmj}}{\kj+\kmj}\cj
\end{equation}
is the single parameter for the $j\ith$ symmetric bin.

The analysis procedure is slightly different in the case of symmetrized binning 
of the~$\bdbpp$ Dalitz plot.  Flavour-specific~$D$ meson decays are not 
needed.  A combined time-dependent fit of the~$\bdbpp$ with~$D$ meson decays 
into~$\cpconj$-specific and~$\kspp$ final states should be performed in order 
to measure the dilution factors $\djth$ 
together with the angle~$\pphi$.  The $\kspp$ final state is still necessary 
since the $\cpconj$-specific final states provide $\mcm$ constraints 
while there are $\mcm+1$ unknown parameters.\footnote{
The continuous ambiguity defined in~Eq.~(\ref{eq:scale}) occurs for the case of 
symmetrized Dalitz plot binning too.  In this case, instead of the phase 
parameters~$\cj$ and~$\sj$, the dilution factors $\djth$ should be scaled.} 

The symmetrization of binning leads to a certain loss of information.  
Particularly, the~$\bdpp$ with~$\cpconj$-specific~$D$ meson decays are not 
sensitive to the~$\cosdbeta$ (Eq.~(\ref{eq:cp_dil_pdf})) in this 
case.  A quantitative evaluation of the sensitivity decline related 
to the symmetrized $\bdbpp$ Dalitz plot partitioning is described in 
the next section.

\section{Feasibility study}\label{sec:feas}
Sensitivity of the described method is assessed with a series of 
toy Monte Carlo (MC) experiments.  The equal-phase~$\dnkpp$ decay Dalitz plot 
binning deduced from the decay model published in~Ref.~\cite{kspp_anton} is 
used.  The values of parameters $\ki$, $\ci$ and $\si$ for that binning are 
taken from measurement in~Ref.~\cite{CLEO_phasees}.

A model-independent measurement of the angle~$\pphi$ in $\bdh$ decays is 
considered as a reference procedure.  The coefficients~$\ccoef$ and~$\scoef$ 
from~Eqs.~(\ref{eq:binned_pdf}) and~(\ref{eq:double_binned_pdf}) for the 
case of~$\bdh$ decays can be obtained using the formal substitutions
\begin{equation}
 \kj \to \frac{1}{2\mcm},\quad \sj\to0,\quad \cj\to\xi_{h^0}^{\cpconj}\lbr -1 \rbr^L,
\end{equation}
where~$\xi_{h^0}^{\cpconj}$ is the~$\cpconj$ eigenvalue of~$h^0$ meson and 
$L$ is the angular moment of $Dh^0$ system.

The MC events are generated with probability density functions (PDFs) of 
the form
\begin{equation}
 p\lbr\dt\rbr = \lbr 1 - f\subbkg\rbr\int\limits_{-\infty}^{\infty} 
 p_{\textrm{true}}^{\textrm{w}}\lbr\dt{}^{\prime}\rbr \mcr\lbr\dt-\dt{}^{\prime}\rbr 
 \mathrm{d}\dt{}^{\prime} + f\subbkg\mcr\lbr\dt\rbr\mathrm{d}\dt{}^{\prime},
\end{equation}
where the resolution function~$\mcr$, employed also as the background PDF, is a 
Gaussian with zero mean and $f\subbkg$~is the background fraction.
The function~$p_{\textrm{true}}^{\textrm{w}}$ is a PDF from~Sec.~\ref{sec:bins} with 
the wrong $B$ meson flavor tagging probability $w$ factor
\begin{equation}\label{eq:raw_pdf_wtag}
  p_{\textrm{true}}^{\textrm{w}}(\dt) \propto
  \bexp\left[1 + q_B\lbr 1-2w \rbr\lbr\ccoef\cos{\dmdt}-\scoef\sin{\dmdt}\rbr\right].
\end{equation}
The tagging power $\etag\equiv \lbr 1 - 2w\rbr^2$ characterizes effective reduction of 
data sample due to non-ideality of a $B$ meson flavour tagging procedure.  The 
tagging power~$\etag=0.3$, typical for $B$ factory experiments, is employed for the~$\belle$ 
and~$\belleii$ and~$\etag=0.08$ is employed for the~$\lhcb$ 
taking into account the recent progress in the flavour-tagging 
algorithms at hadronic machines~\cite{bdd_lhcb_tag}.\footnote{The flavor tagging power
$\etag$ at $\lhcb$ strongly depends on the decay channel and the actual value 
for the~$\bdpp$ decay may differ from the adopted in this work value $0.08$.}
The values of PDF parameters for the $\bellebii$ and $\lhcb$ are 
chosen based on results from 
Refs.~\cite{cosbeta_belle, sinbeta_bcud, bdpp_lhcb} and are shown 
in~Table~\ref{tab:conditions}.

\begin{table}
 \centering
 \caption{Experimental conditions adopted in numerical experiments.}
 \label{tab:conditions}
 \begin{tabular}{lcc}
 \hline\hline
 Parameter                         & $\belle$ \& $\belleii$ & $\lhcb$ \\ \hline
 Time resolution $\sigma_t$ (ps)   & $1.25$                 & $0.06$ \\
 Tagging power $\etag$\ (\%)       & $30$                   & $8$ \\
 Background fraction (\%)          & $30$                   & $5$ \\
 \hline\hline
 \end{tabular}
\end{table}

Table~\ref{tab:yield} shows estimates of the signal yields for the $\belle$, 
$\belleii$ and~$\lhcb$ experiments.  The estimates for $\belle$ are obtained 
using the results from Refs.~\cite{bdpp_belle,cosbeta_belle,sinbeta_bcud}.  The 
estimates for $\belleii$ are obtained by extrapolating the $\belle$ 
yields assuming the same experimental conditions and~$50$ times larger 
integrated luminosity.  The estimate signal yields corresponding to the 
data collected by $\lhcb$ in $2010$~--~$2012$  are based on the results from 
Refs.~\cite{bdpp_lhcb,bdk_lhcb,bdk_kspp_lhcb}.  This period of data taking is 
referred to as~$\runi$.  The estimates for the $\lhcb$ signal yields 
corresponding to the end of current data taking period ($\runii$) 
and to the period of data taking after the planned upgrade ($\upg$)~\cite{lhcb_upgr}
are roughly estimated to be, respectively, $4$ and $70$ times larger 
than the~$\runi$ values, assuming the corresponding luminosity integrals 
equal~$8~\ifb$ and $50~\ifb$.

\begin{table}
 \centering
 \caption{Estimates of the signal yields for the~$\bn\to \dnbar\{h^0,\ \pi^+\pi^-\}$, 
 $\dnbar\to\{f_{\cpconj},\ \kspp\}$ (and $\cconj$-conjugated) decays at 
 the~$\belle$, $\belleii$ and $\lhcb$ experiments.}
 \label{tab:yield}
\begin{tabular}{lccccc}
 \hline\hline
\multirow{2}{*}{Mode} & \multirow{2}{*}{$\belle$}  & \multirow{2}{*}{$\belleii$} 
             & \multicolumn{3}{c}{$\lhcb$} \\
             &   & & $\runi$ & $\runii$ & $\upg$ \\
 \hline
 $\bdcppp$    & $1.0\cdot10^3$ & $50 \cdot10^3$ & $2.0\cdot10^3$ & $8\cdot10^3$ & $140\cdot10^3$ \\
 $\bdkspppp$  & $1.3\cdot10^3$ & $65 \cdot10^3$ & $1.2\cdot10^3$ & $5\cdot10^3$ & $84 \cdot10^3$ \\
 $\bdcph$     & $0.8\cdot10^3$ & $40 \cdot10^3$ & --- & --- & --- \\
 $\bdkspph$   & $1.0\cdot10^3$ & $50 \cdot10^3$ & --- & --- & --- \\
 \hline\hline
 \end{tabular}
\end{table}

The signal yields for $\bdpp$ with flavour-specific $D$ 
meson decays are relatively large for both $\belle$ and~$\lhcb$.  Thus, 
the uncertainties related to the parameters $k_j$ are neglected.

\subsection{Parameters of the \texorpdfstring{$\bdbpp$}{B0 -> anti-D0 pi+ pi-} 
decay binned Dalitz plot} \label{sec:model} 
\begin{figure}
 \centering
\subfloat[]{\label{fig:dpp_dp}    \includegraphics[width=0.32\textwidth]{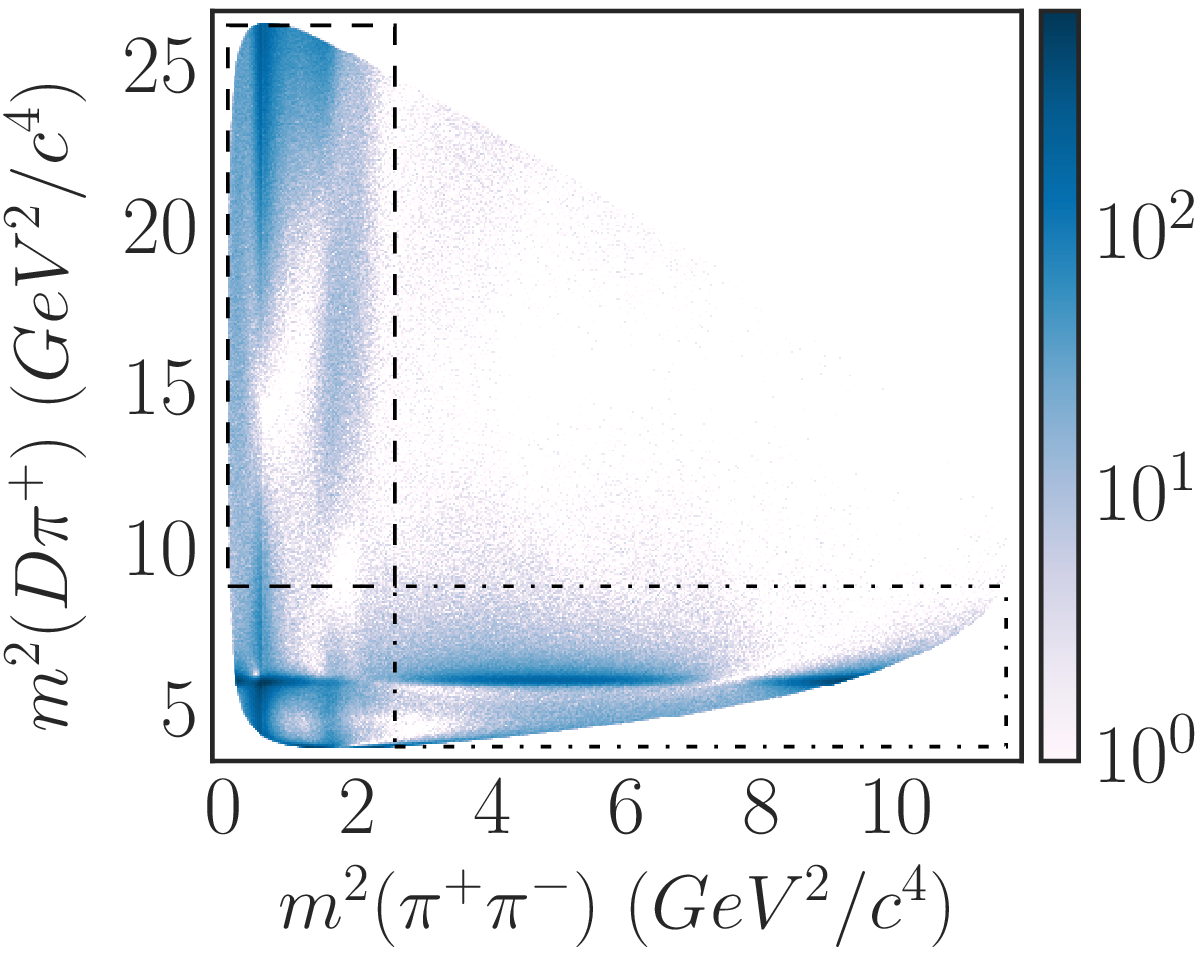}}
\subfloat[]{\label{fig:dpp_mdpip} \includegraphics[width=0.32\textwidth]{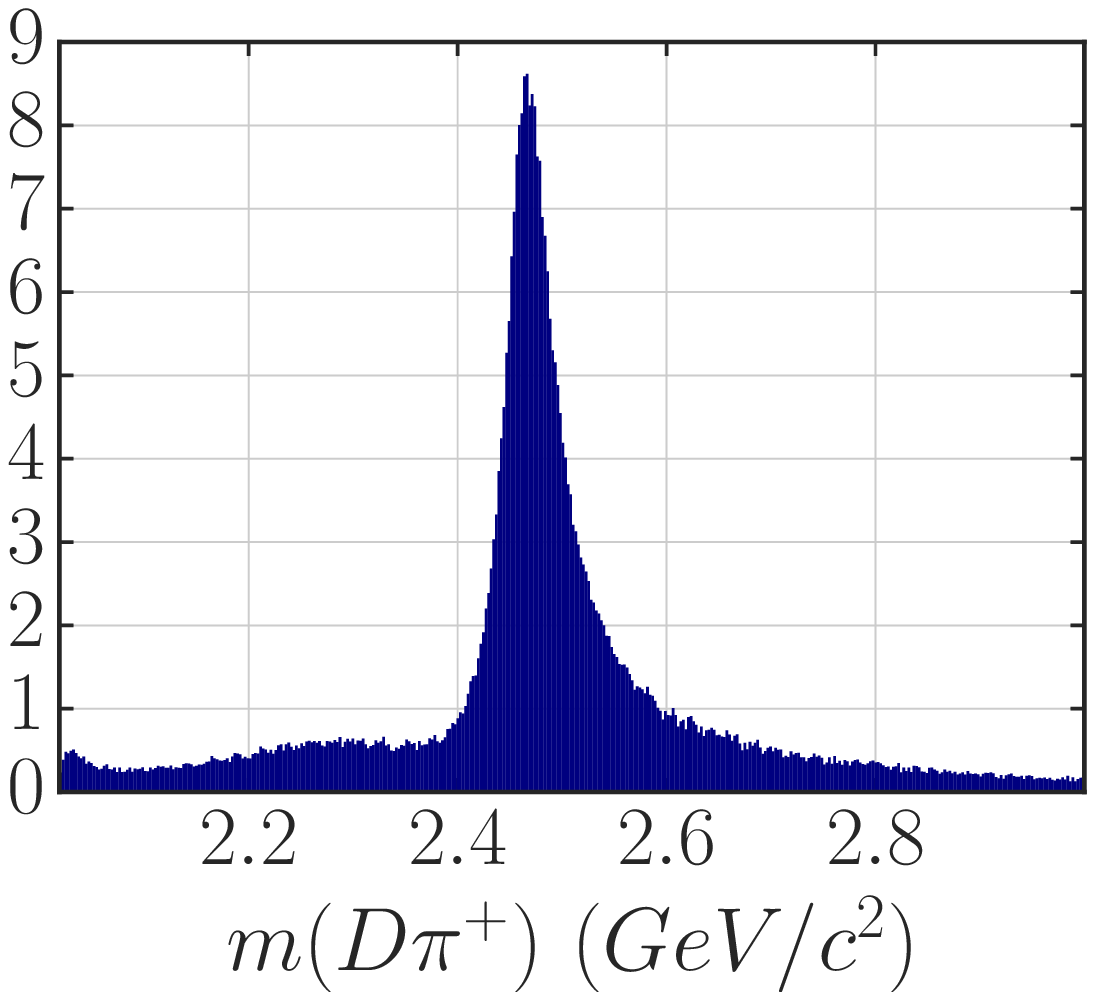}}
\subfloat[]{\label{fig:dpp_mpipi} \includegraphics[width=0.32\textwidth]{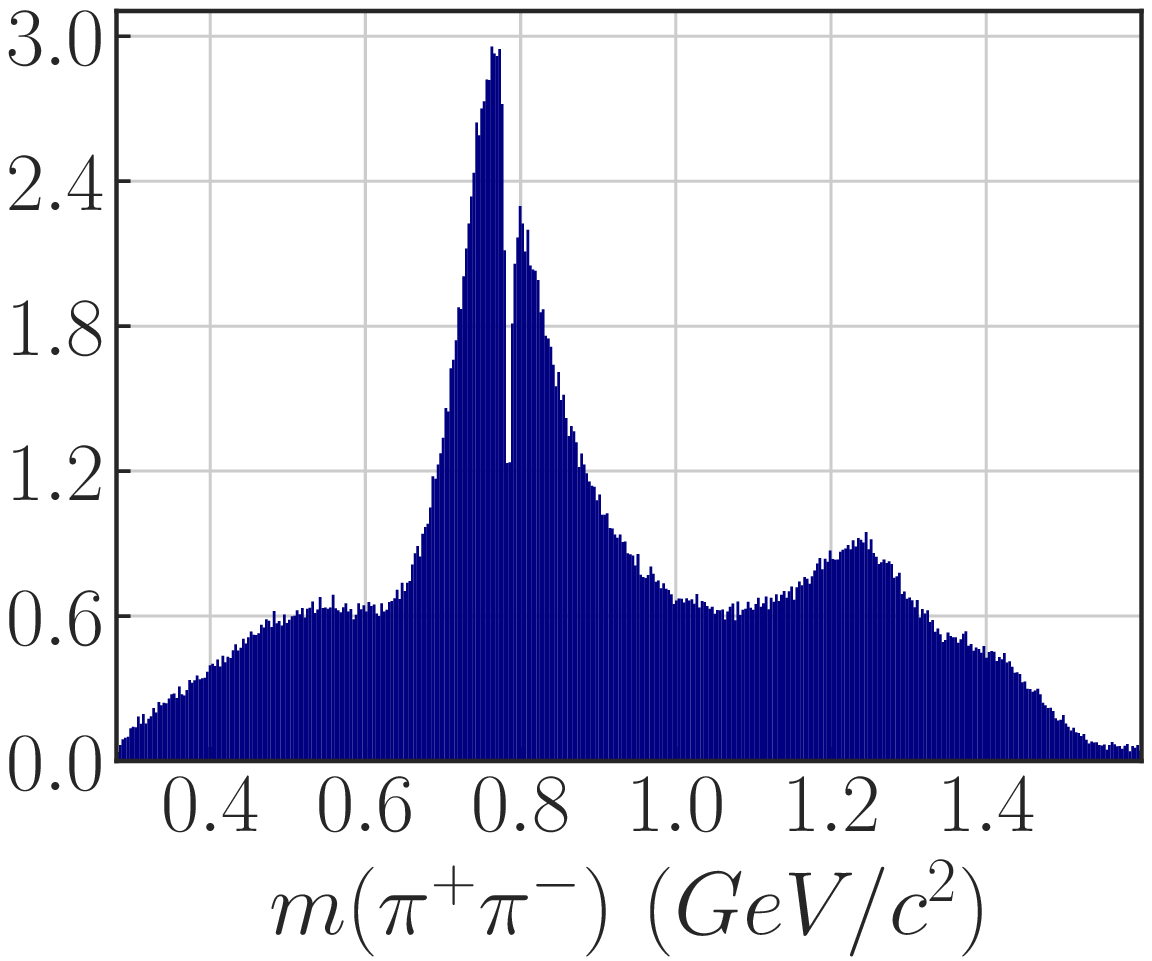}}
\caption{Dalitz plot distribution of the~$\bdbpp$ decay~(a), 
$m\lbr\dn\pi^+\rbr$ distribution below~$3~\gevcsq$ for 
$m\lbr\pipi\rbr>1.6~\gevcsq$~(b), and $m\lbr\pipi\rbr$ distribution 
below $1.6~\gevcsq$ for $m\lbr\dn\pi^+\rbr>3~\gevcsq$~(c).  The distributions 
are obtained with the $\bdbpp$ decay amplitude model described in 
Appendix~\ref{app:decay_ampl}.  
The dashed and dot-dashed regions on the Dalitz plot correspond to the 
distributions on the subplots (b) and (c), respectively.}
\label{fig:unbinned_bdpp}
\end{figure}
Two models of the $\bdbpp$ decay amplitude are available 
in~Refs.~\cite{bdpp_belle,bdpp_lhcb}.  A~simplified version of the model from 
Ref.~\cite{bdpp_belle} is used in this study (see Appendix~\ref{app:decay_ampl}).  
The Dalitz distribution and distributions of 
the~$\dn\pi^+$ and~$\pi^+\pi^-$ invariant masses obtained with this model 
are shown in Figure~\ref{fig:unbinned_bdpp}.

The equal-phase binning of the $\bdbpp$ decay Dalitz plot into $16$ bins is 
performed using this model.  The bin regions obtained and corresponding 
values of the parameters $\kj$, $\cj$ and $\sj$ are shown 
in~Figure~\ref{fig:binned_bdpp}.

\begin{figure}
 \centering
\subfloat[]{\label{fig:dpp_bin2} \includegraphics[width=0.335\textwidth]{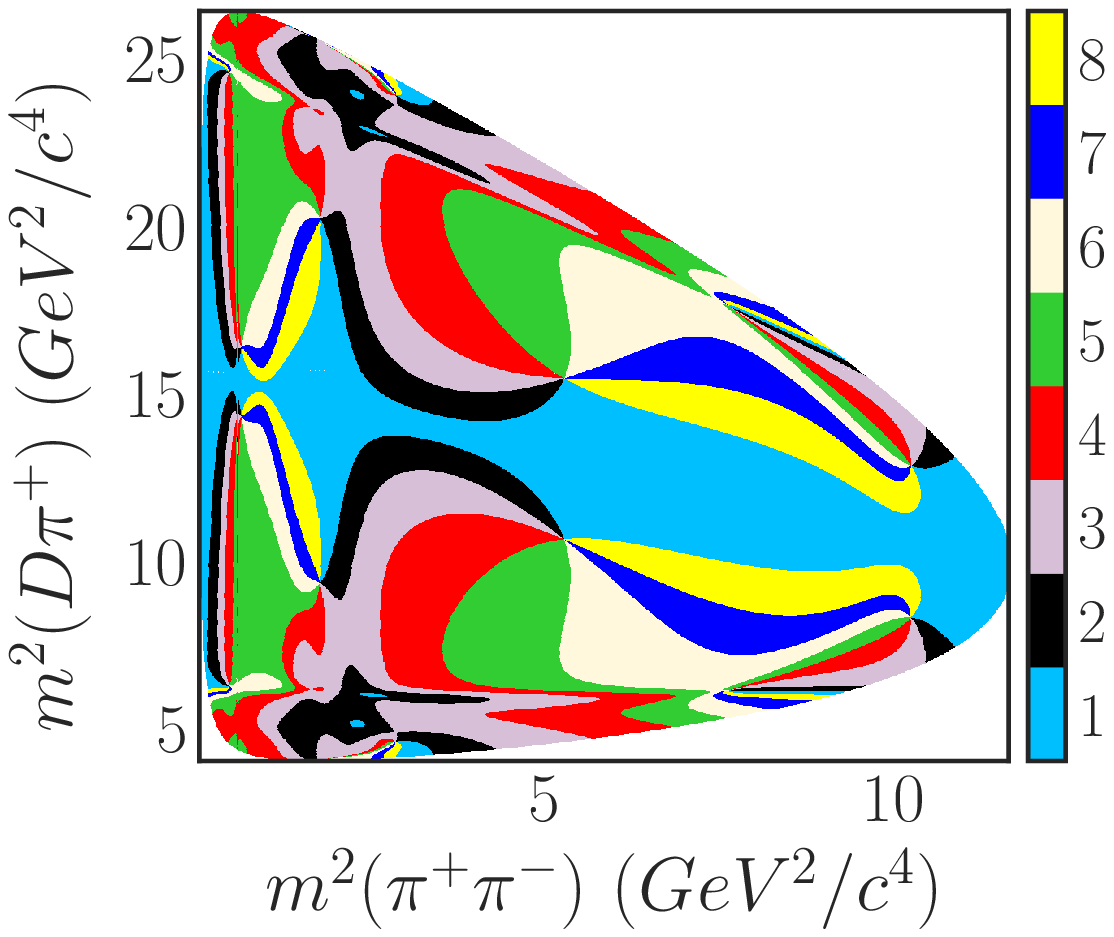}}%
  \hspace{0.01 cm}%
\subfloat[]{\label{fig:dpp_k}    \includegraphics[width=0.335\textwidth]{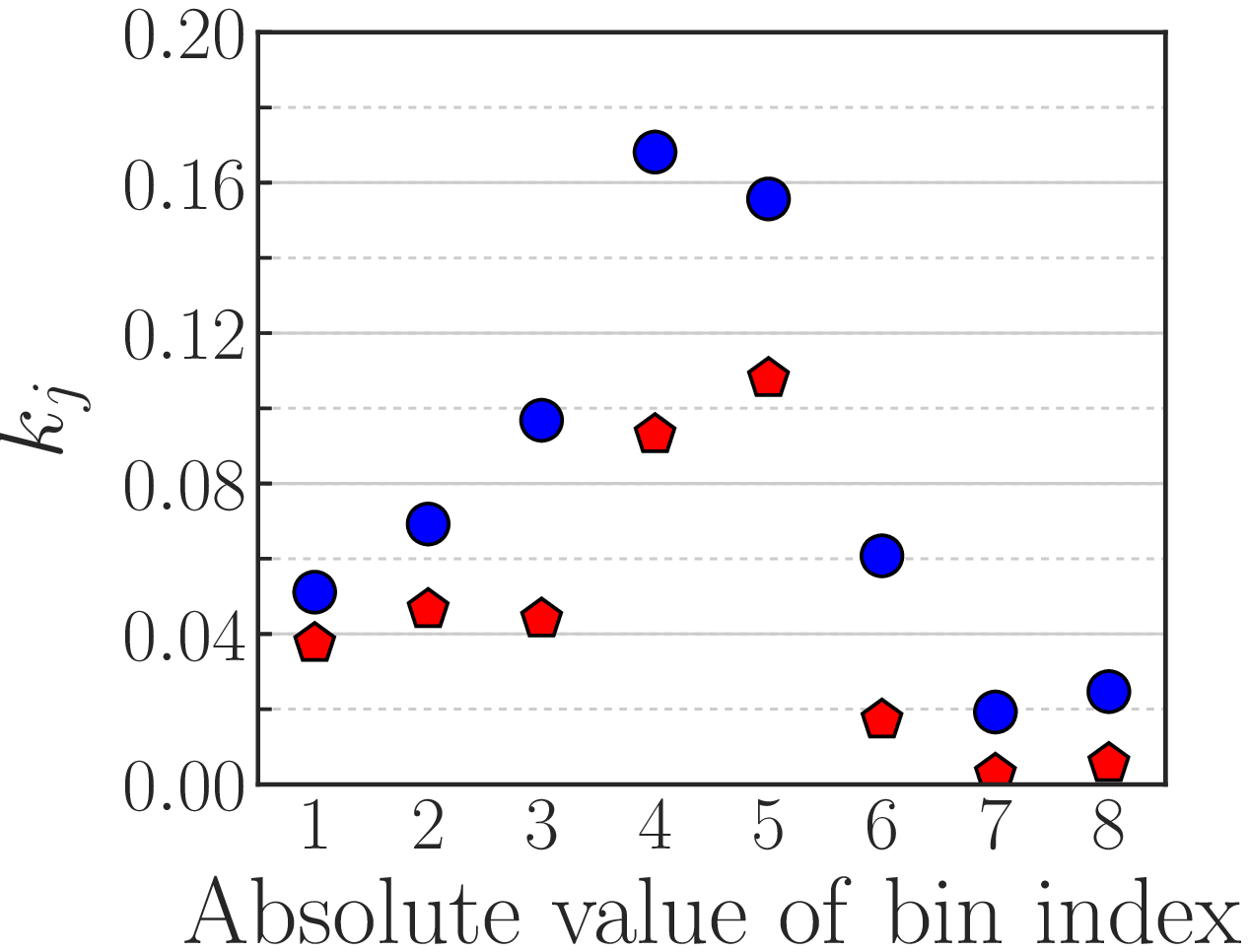}}%
  \hspace{0.01 cm}%
\subfloat[]{\label{fig:dpp_cs}   \includegraphics[width=0.29\textwidth]{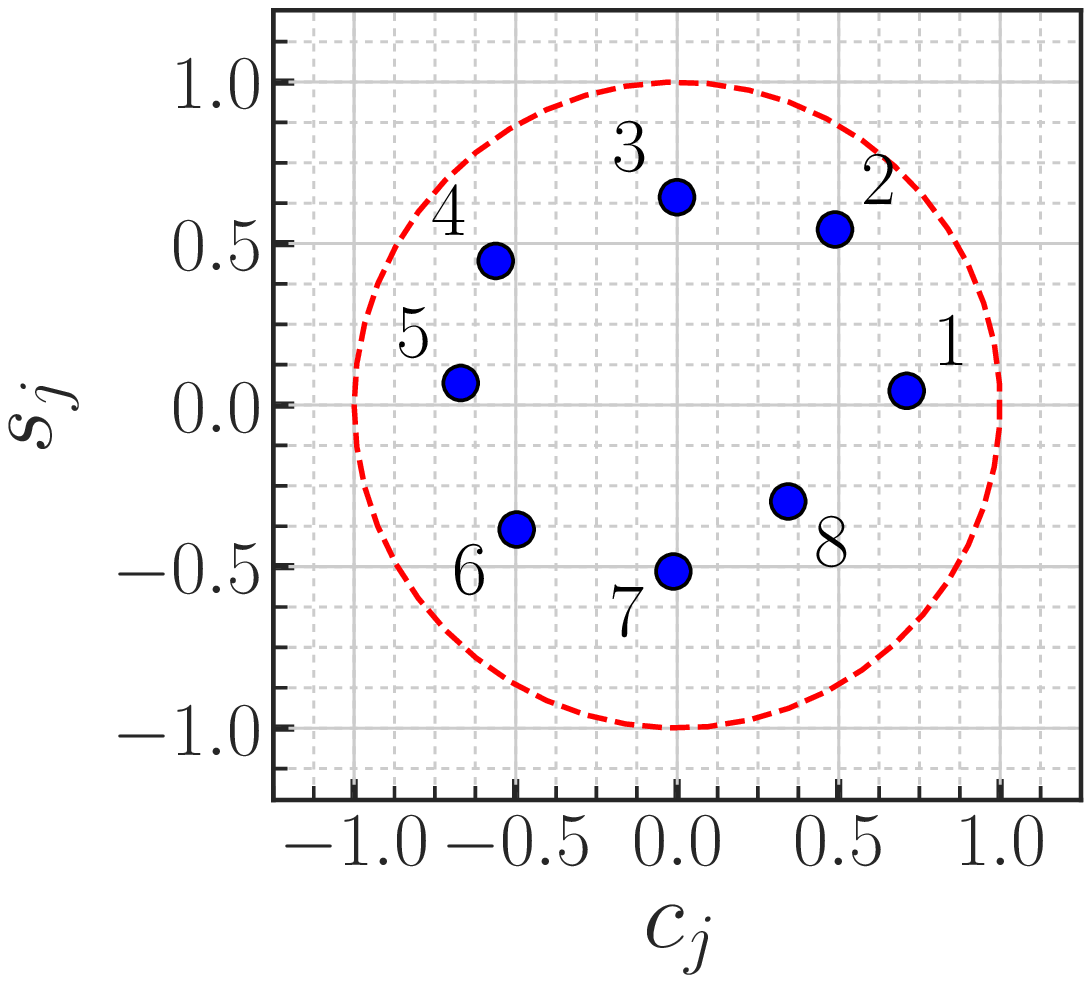}}%
\caption{Equal-phase Dalitz plot binning~(a), values of the parameters $\kj$ 
for $j>0$ (blue circles) and $j<0$ (red pentagons)~(b), and values of the 
parameters $\cj$ and $\sj$ (blue circles)~(c) obtained with the $\bdbpp$ decay 
amplitude model described in Appendix~\ref{app:decay_ampl}.}
\label{fig:binned_bdpp}
\end{figure}

\subsection{Numerical experiments}\label{sec:toymc}
Three approaches to measure the angle~$\pphi$ are considered.  
Each approach implies the joint analysis of~$\dt$ distributions 
for the~$\bdpp$ with $D$ meson decays into $\cpconj$-specific 
and~$\kspp$ final states.  These approaches are:
\begin{enumerate}
 \item The fit based on Eqs.~(\ref{eq:cp_pdf}) and~(\ref{eq:dd_pdf}) 
 with~$17$ free parameters: eight~$\lbr \cj, \sj\rbr$ pairs and the 
 angle~$\pphi$.
 \item The fit using symmetrized $\bdpp$ decay Dalitz plot binning with 
 nine free parameters: eight dilution factors~$\djth$, defined 
 in~Eq.~(\ref{eq:dilut}), and the angle~$\pphi$.
 \item Model-independent measurement of the angle~$\pphi$ 
 in the~$\bdh$ decays as a reference.  The angle~$\pphi$ is the only free 
 parameter in this case.
\end{enumerate}

\begin{table}
\centering
\caption{Estimates of the angle~$\pphi$ measurement statistical precision for 
the three schemes with the input value $\pphi=22\grad$.
}
\label{tab:precision}
\begin{tabular}{lrrrrr}
 \hline\hline
 \multirow{2}{*}{Measuring scheme} &
 \multirow{2}{*}{$\belle$}  & 
 \multirow{2}{*}{$\belleii$} &
 \multicolumn{3}{c}{$\lhcb$} \\
                                  &           &            & $\runi$   & $\runii$ & $\upg$ \\
 \hline
 $\bdpp$                          & $\approx10\grad$ & $1.5\grad$ & $\approx15\grad$ & $ 6\grad$ & $1.5\grad$ \\
 $\quad$ Only $\dkpp$             & $\approx15\grad$ & $2  \grad$ & $\approx20\grad$ & $ 7\grad$ & $2  \grad$ \\
 \hline
 $\bdpp$ (symm)                   & $\approx15\grad$ & $2  \grad$ & $\approx20\grad$ & $10\grad$ & $2  \grad$ \\
 $\quad$ Only $\dkpp$             & $\approx20\grad$ & $2.5\grad$ & $\approx25\grad$ & $13\grad$ & $3  \grad$ \\
 \hline
 $\bdsth$                         & $ 5\grad$ & $0.7\grad$ & --- & --- & --- \\
 $\quad$ Only $\dkpp$             & $ 7\grad$ & $1.1\grad$ & --- & --- & --- \\
 $\quad$ Only $D\to f_{\cpconj}$  & $ 6\grad$ & $0.8\grad$ & --- & --- & --- \\
 \hline\hline
 \end{tabular}
\end{table}

The statistical precision of the angle $\pphi$ measurement for the initial 
value~$\pphi=22\grad$, obtained with each of the three approaches, 
is shown in~Table~\ref{tab:precision}.  The analysis of~$\bdpp$ decays 
provides precision about $1.5$ times worse than the analysis of~$\bdh$ 
decays.  The prospects for the analysis of $\bdh$ decays at~$\lhcb$ are 
not clear since there are neutral particles in the final state. 
The~$\belleii$ and upgraded $\lhcb$ have comparable potential to measure 
the angle~$\pphi$ in~$\bdpp$ decays.  A combination of the results from~$\bdh$ 
and $\bdpp$ analyses would yield the~$\pphi$ precision in $\btocud$ 
transitions below one degree.\footnote{At the moment, the  
uncertainty related to the~$\ci$ and~$\si$ parameters measurement 
is about~$1.1\grad$, as it is stated in Ref.~\cite{cosbeta_belle}.  
The precision level below one degree can be achieved only if a more 
precise measurement of the parameters~$\ci$ and~$\si$ appears.  
Such a measurement can be provided by the BESIII collaboration and by a future 
Super $c$-$\tau$ factory experiment.
}

\begin{figure}
 \centering
 \subfloat[]{\label{fig:dil_fit}
  \includegraphics[width=0.3\textwidth]{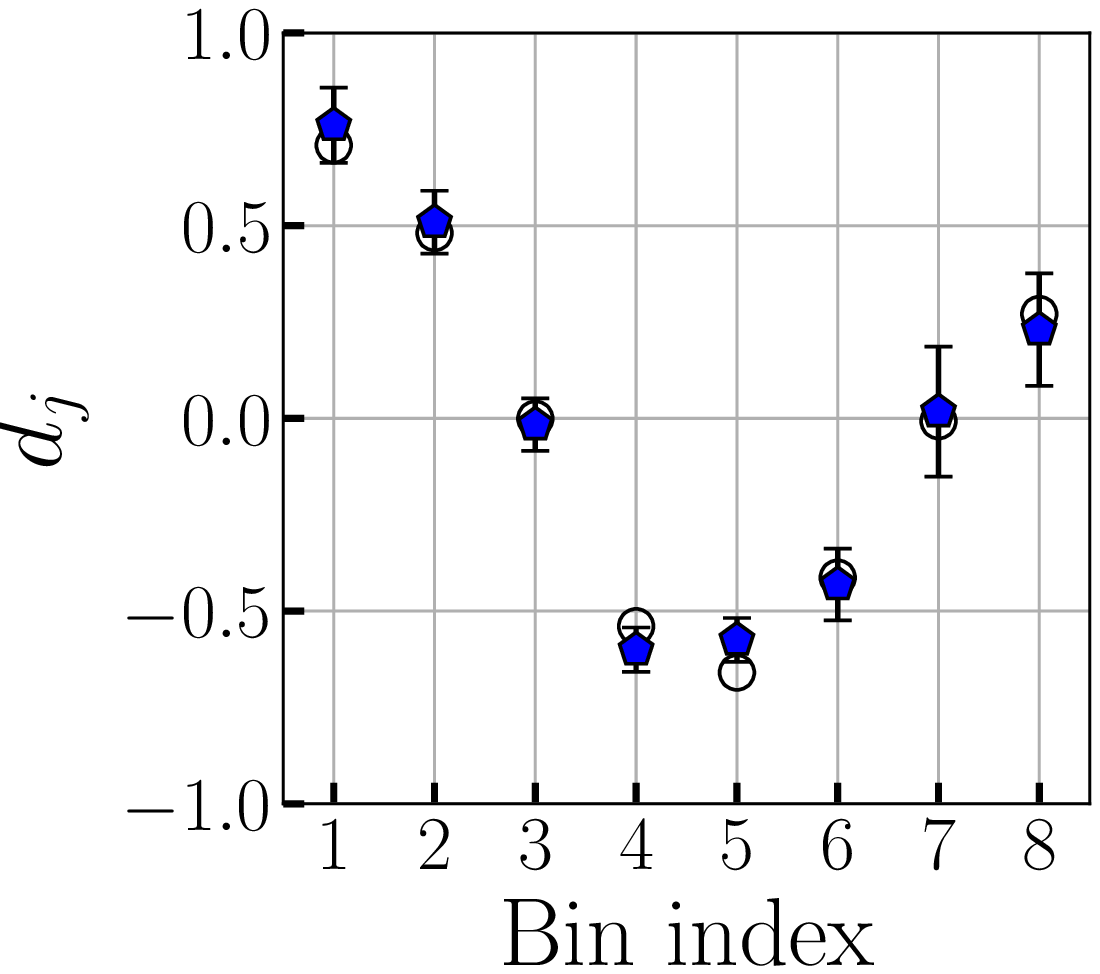}}
  \hspace{0.5 cm}
 \subfloat[]{\label{fig:csfit}
  \includegraphics[width=0.3\textwidth]{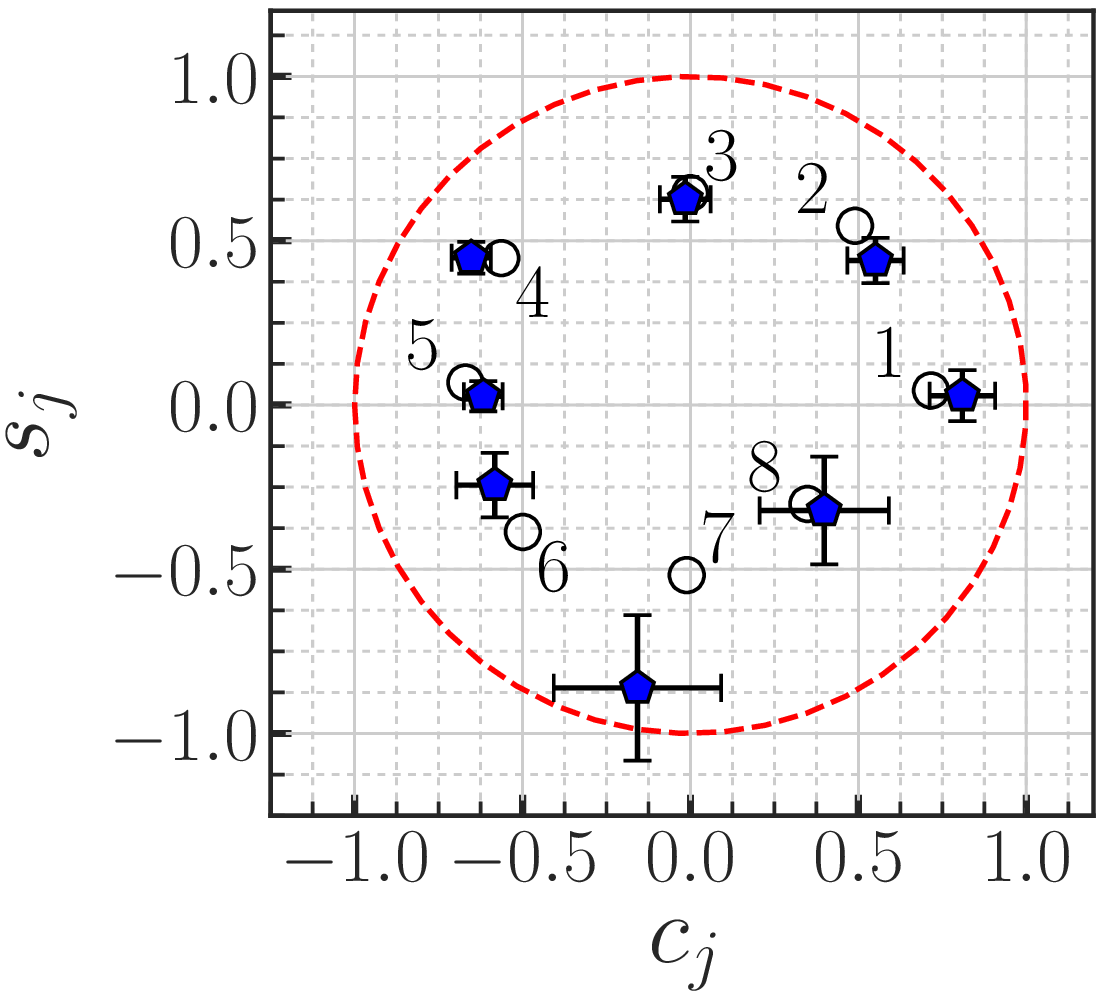}}
\caption{Results of MC simulation: dilution factors $\djth$~(a) and phase 
parameters~$\cj$ and~$\sj$~(b).  Empty circles show the input values, blue 
pentagons with error bars show the fit results obtained for the expected 
$\belleii$ statistics and experimental conditions.}
\label{fig:phase_fit}
\end{figure}

Figure~\ref{fig:phase_fit} illustrates prospects for the $\belleii$ experiment:  
a fit result for the dilution factors~$\djth$ (Figure~\ref{fig:dil_fit}) 
and for the parameters $\cj$ and $\sj$ (Figure~\ref{fig:csfit}) 
obtained with MC simulation for the input value~$\pphi=22\grad$.

The results presented are obtained with a simple method of the Dalitz plot 
binning (the equal-phase binning).  It is shown in~Refs.~\cite{BP_phi3_model2, 
CLEO_phasees} that the binning can be optimized to improve the statistical 
sensitivity by a factor of about~$1.2$.

\section{Conclusions}
A novel model-independent approach to measure the~$\ckm$ angle~$\pphi$ with 
time-dependent analysis of the~$\bdpp$ decays dominated by the tree quark 
transition is proposed.  It is shown that the angle~$\pphi$ and the parameters 
of binned~$\bdbpp$ decay Dalitz plot can be obtained from the single 
measurement.  Statistical precision of the method is comparable to that of 
the model-independent angle~$\pphi$ measurement in~$\bdh$ decays.

The fact that only charged particles compose the final states of~$\bdpp$, 
$D\to\fcp$ and~$\dkpp$ decay chains for such $\fcp$ as~$K^+K^-$, $\pi^+\pi^-$, 
and $\phi\ks$ provides good experimental perspectives for $\lhcb$.

The angle~$\pphi$ can be measured with the one-degree precision level at the 
$\belleii$ and $\lhcb$ experiments in $\btocud$ transitions in a 
model-independent way, namely without the need to model neither the 
$\dnkpp$ nor the $\bdbpp$ decay amplitudes. The combined 
time-dependent analysis of~$\bdh$ and~$\bdpp$ decays with $D$ meson decaying into a $\fcp$ 
($\fcp=K^+K^-$, $\ks\pin$ etc.) and $\kspp$ states should be performed 
in order to achieve such precision.

The measurement bias inherent in the proposed method due to the neglect of 
the suppressed transition $\btoucd$ and charm mixing is of order of $0.2\grad$ 
(see Appendix~\ref{app:systematics-evaluation}) and can be considered 
as a probably non-dominant systematic uncertainty.

\acknowledgments
Authors would like to thank Anton Poluektov and Simon Eidelman for 
useful discussions. Work of V.V. was supported by the Grant of the Russian 
Federation Government, Agreement \# 14.W03.31.0026 from 15.02.2018.

\appendix
\section{\boldmath The \texorpdfstring{$\bdbpp$}{B0 -> anti-D0 pi+ pi-} decay amplitude model}\label{app:decay_ampl}
A simple isobar model of the $\bdbpp$ decay amplitude, inspired by the result 
from Ref.~\cite{bdpp_belle}, is used in numerical experiments.  The resonances 
constituting the model are listed in~Table~\ref{tab:bdpp_res_list}.  Each 
resonance is described by a relativistic Breit-Wigner function~\cite{breit-wigner}.  
Energy-dependent resonance width and Blatt-Weisskopf barrier factors~\cite{blatt-weiisskopf1,
blatt-weiisskopf2} are used.

\begin{table}[t]
 \caption{List of resonances included in the $\bdbpp$ decay amplitude model.  
The resonance fit fraction is denoted by $\mcf$ and the resonance amplitude phase 
is denoted by $\varphi$}
  \label{tab:bdpp_res_list}
  \centering
 \begin{tabular}{lrrrrr}
  \hline\hline
   \multicolumn{1}{c}{Name} &
   \multicolumn{1}{c}{$M$ ($\gevcsq$)} &
   \multicolumn{1}{c}{$\Gamma$ ($\mev$)} &
   \multicolumn{1}{c}{$J$} &
   \multicolumn{1}{c}{$\mcf$ (\%)} &
   \multicolumn{1}{c}{$\varphi$ ($\mathrm{deg}$)}
   \\ \hline
   $D^*_2(2460)$ & $2.4657$   & $49.6$      & $2$  & $29.9$ & $0$      \\
   $D^*_v$       & $2.01$     & $10^{-4}$   & $1$  & $7.6$  & $-145.0$ \\
   $D^*_0(2400)$ & $2.308$    & $276.11$    & $0$  & $6.5$  & $-165.0$ \\ \hline
   $\rho^0(770)$ & $0.7756$   & $144$       & $1$  & $36.3$ & $103.7$  \\
   $\omega(782)$ & $0.7826$   & $8.49$      & $1$  & $0.5$  & $-88.4$  \\
   $\rho(1450)$  & $1.465$    & $310$       & $1$  & $0.4$  & $-76.3$  \\
   $f_2(1270)$   & $1.275$    & $185$       & $2$  & $7.5$  & $-97.6$  \\
   $f_0(500)$    & $0.513$    & $335$       & $0$  & $10.0$ & $80.8$   \\
   $f_0(1370)$   & $1.434$    & $173$       & $0$  & $1.8$  & $-139.2$ \\
  \hline\hline
 \end{tabular}
\end{table}

The model describes two main channels $\bn\to\dn\rho^0(770)$ and 
$\bn\to D_2^{\ast}(2460)\pi$.  The scalar $D^*_0(2400)$ and virtual vector $D^*_v$ 
resonances describe the remaining $\dn\pi$ structure.  Following Ref.~\cite{bdpp_belle} 
we call $D^*_v$ virtual since the veto $\left|m(D\pi) - m(D^*)\right| > 3~\mevcsq$ 
is imposed and only the tail of $D^*_v$ resonance contributes the amplitude.  

The remaining $\pipi$ structure is described by the wide scalar $f_0(500)$, narrow vector~$\omega$ 
interfering destructively with $\rho^0(770)$ and resonances $\rho(1450)$, $f_2(1270)$ and $f_0(1370)$
responsible for the $\pipi$ mass spectrum above $1~\gevcsq$.  

\section{\boldmath Formalism accounting for the \texorpdfstring{$\btoucd$}{b -> c ubar d} 
transition}\label{app:bucbard}
A precise measurement of the angle $\pphi$ in the $\btocud$ transitions requires 
understanding the bias due to the neglect of the suppressed decay $\bdnpp$ and 
charm mixing.  Both processes produce additional interfering amplitudes 
for the $B^0\to \dnbar \pi^+\pi^-$, $\dbkpp$ decay shown on the scheme at
Figure~\ref{fig:suppressed-transitions}.

\begin{figure}
\centering
  \includegraphics[width=0.5\textwidth]{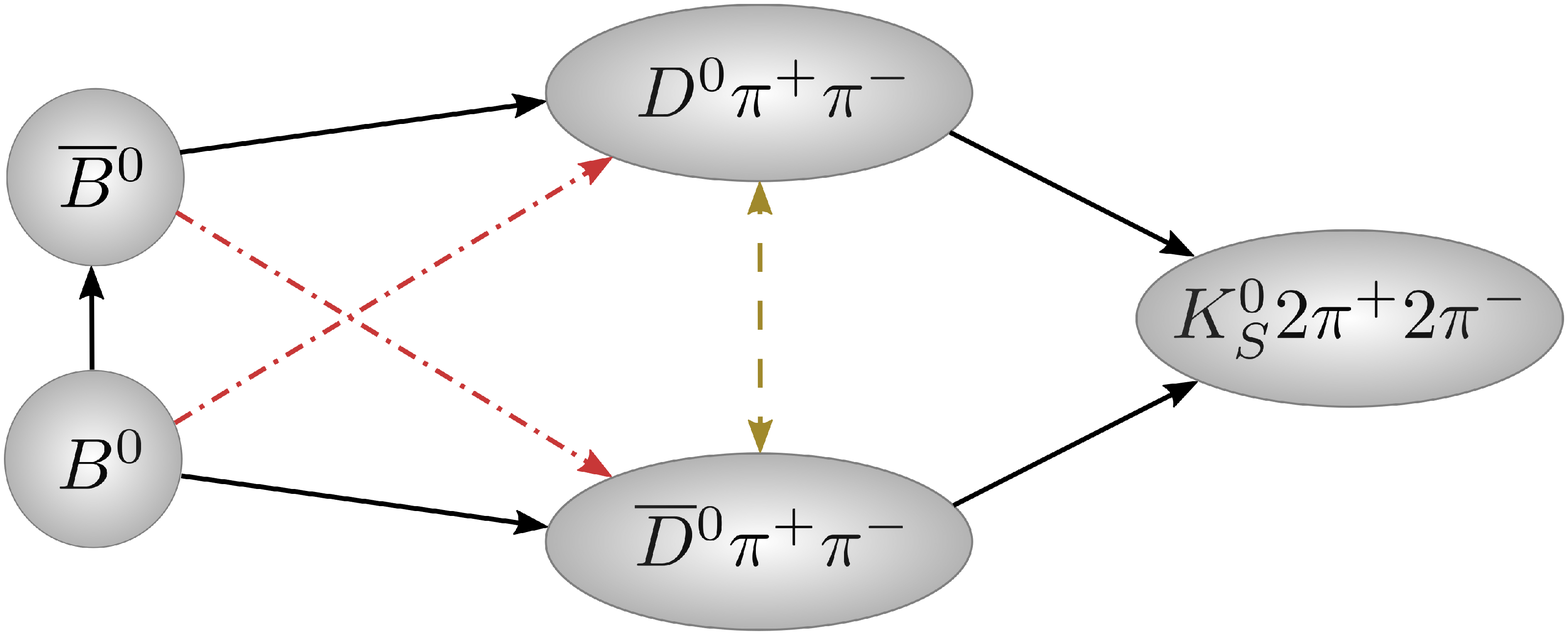}
 \caption{Transitions scheme of the $\bdpp$, $\dkpp$ process. Black solid arrows
 denote dominant transitions, red dash-dotted arrows denote suppressed $B\to D$ and 
 $\bbar\to\dbar$ transitions, brown dashed arrows denote $\dn$-$\dnbar$ oscillations.}
 \label{fig:suppressed-transitions}
\end{figure}

This appendix extends the formalism presented in sections~\ref{sec:pdfs}~and~\ref{sec:bins}
and accounts for the $\bdnpp$ decay.  Corrections due to the charm mixing are 
considered in appendix~\ref{app:charm-mixing}.  Quantitative estimates of the bias 
due to the neglect of these processes are described in appendix~\ref{app:systematics-evaluation}.

The $\bn\to D\pi^+\pi^-$, $\dkpp$ decay amplitude including the $\btoucd$ transition 
(without charm mixing) reads
\begin{equation}\label{eq:decay_amp_scs}
\begin{split}
 \mca_{B\to f}&\dtbvardvar\\
 &= \abdbar\bvar \adbar\dvar \cos\hdmdt \\
 &+ \abd\bvar\ad\dvar e^{i\gphi} \cos\hdmdt \\
 &+ i\abbard\bvar \ad\dvar e^{-2i\pphi} \sin\hdmdt \\
 &+ i\abbardbar\bvar \adbar\dvar e^{-i\lbr2\pphi + \gphi\rbr} \sin\hdmdt,
\end{split}
\end{equation}
where $\gphi$ is the $\ckm$ phase.  The corresponding 
$\bnbar\to D\pi^+\pi^-$, $\dkpp$ decay amplitude is
\begin{equation}\label{eq:decay_bar_amp_scs}
\begin{split}
 \mca_{\bbar\to f}&\dtbvardvar\\
 &=  \abbard   \bvar \ad   \dvar \cos\hdmdt \\
 &+  \abbardbar\bvar \adbar\dvar e^{-i\gphi} \cos\hdmdt \\
 &+ i\abdbar   \bvar \adbar\dvar e^{2i\pphi} \sin\hdmdt \\
 &+ i\abd      \bvar \ad   \dvar e^{i\lbr2\pphi + \gphi\rbr} \sin\hdmdt,
\end{split}
\end{equation}

The decay probability densities corresponding to the 
amplitudes~(\ref{eq:decay_amp_scs}) and~(\ref{eq:decay_bar_amp_scs}) are 
\begin{equation}\label{eq:time-pdf-udf}
 p\dtbvardvar = \ucoef + q_B\left[\ccoef\cos{\dmdt} + \scoef\sin{\dmdt}\right],
\end{equation}
where
\begin{equation}\label{eq:ucoef_complete}
 \begin{split}
  \ucoef &= \frac{1}{2} p_D\dvar    \left[ \pbd   \bvar + \pbdbar\bvarinv \right] \\
         &+ \frac{1}{2} p_D\dvarinv \left[ \pbdbar\bvar + \pbd   \bvarinv \right] \\
         &+ \sqrt{\pbdbar\bvar    \pbd\bvar    p_D\dvar p_D\dvarinv} \\
         &\times \cos{\lbr\dbd + \gphi - \deld\rbr} \\
         &+ \sqrt{\pbdbar\bvarinv \pbd\bvarinv p_D\dvar p_D\dvarinv} \\
         &\times \cos{\lbr\dbd - \gphi + \deld + \psibb - \psib\rbr},
 \end{split}
\end{equation}
\begin{equation}\label{eq:ccoef_complete}
 \begin{split}
  \ccoef &= \frac{1}{2} p_D\dvar    \left[ \pbd   \bvar - \pbdbar\bvarinv \right] \\
         &+ \frac{1}{2} p_D\dvarinv \left[ \pbdbar\bvar - \pbd   \bvarinv \right] \\
         &+ \sqrt{\pbdbar\bvar    \pbd\bvar    p_D\dvar p_D\dvarinv} \\
         &\times \cos{\lbr\dbd + \gphi - \deld\rbr} \\
         &- \sqrt{\pbdbar\bvarinv \pbd\bvarinv p_D\dvar p_D\dvarinv} \\
         &\times \cos{\lbr\dbd - \gphi + \deld + \psibb - \psib\rbr},
 \end{split}
\end{equation}
\begin{equation}\label{eq:scoef_complete}
 \begin{split}
  \scoef = &-\sqrt{\pbdbar\bvar \pbdbar\bvarinv p_D\dvar p_D\dvarinv} \\
           &\times \sin{\lbr\psib - 2\pphi - \deld\rbr} \\
           &-\sqrt{\pbd   \bvar \pbd   \bvarinv p_D\dvar p_D\dvarinv} \\
           &\times \sin{\lbr\psibb - 2\pphi + \deld - 2\gphi\rbr} \\
           &-p_D\dvarinv\sqrt{\pbdbar\bvar    \pbd\bvarinv} \\
           &\times \sin{\lbr\psibb - 2\pphi + \dbd - \gphi\rbr} \\
           &-p_D\dvar   \sqrt{\pbdbar\bvarinv \pbd\bvar} \\
           &\times \sin{\lbr\psib  - 2\pphi - \dbd - \gphi\rbr}.
 \end{split}
\end{equation}

The following notation is used (compare with Eq.~(\ref{eq:phases-definition})):
\begin{subequations}\label{eq:phase_nonation}
 \begin{align}
  \label{eq:psib}
  \psib  &= \mathrm{arg}\lbr\frac{\abdbar\bvarinv}{\abdbar\bvar}\rbr, \\ 
  \label{eq:psibb}
  \psibb &= \mathrm{arg}\lbr\frac{\abd   \bvarinv}{\abd   \bvar}\rbr, \\
  \label{eq:delb}
  \delb  &= \mathrm{arg}\lbr\frac{\abd   \bvar   }{\abdbar\bvar}\rbr, \\
  \deld  &= \mathrm{arg}\lbr \frac{\ad    \dvarinv}{\ad    \dvar} \rbr.
 \end{align}
\end{subequations}

Integration of Eqs.~(\ref{eq:ucoef_complete}, \ref{eq:ccoef_complete}, 
\ref{eq:scoef_complete}) over the $i^{\mathrm{th}}$ bin of $D$ Dalitz 
plot and the $j^{\mathrm{th}}$ bin of $B$ Dalitz plot
leads to
\begin{equation}\label{eq:ucoef_binned_complete}
\begin{split}
 \ucoef_{ij} &= \frac{1}{2}\lbr\kmi\kj + \ki\kmj\rbr \\
             &+ \rbd\sqrt{\ki\kmi\kj\kbj}
             \left[\lbr\ctj \ci + \stj \si\rbr\cos{\gphi} -
                   \lbr\stj \ci - \ctj \si\rbr\sin{\gphi}\right]\\
             &+\rbd\sqrt{\ki\kmi\kmj\kbmj}
             \left[\lbr\ctmj\ci - \stmj\si\rbr\cos{\gphi} +
                   \lbr\stmj\ci + \ctmj\si\rbr\sin{\gphi}\right]\\
             &+\frac{1}{2}\rbd^2\lbr\ki\kbj + \kmi\kbmj\rbr,
\end{split}
\end{equation}
\begin{equation}\label{eq:ccoef_binned_complete}
 \begin{split}
  \ccoef_{ij} &= \frac{1}{2}\lbr\kmi\kj - \ki\kmj\rbr \\
             &+ \rbd\sqrt{\ki\kmi\kj\kbj}
             \left[\lbr\ctj\ci + \stj\si\rbr\cos{\gphi} -
                   \lbr\stj\ci - \ctj\si\rbr\sin{\gphi}\right]\\
             &-\rbd\sqrt{\ki\kmi\kmj\kbmj}
             \left[\lbr\ctmj\ci - \stmj\si\rbr\cos{\gphi} +
                   \lbr\stmj\ci + \ctmj\si\rbr\sin{\gphi}\right]\\
             &+\frac{1}{2}\rbd^2\lbr\ki\kbj - \kmi\kbmj\rbr,
 \end{split}
\end{equation}
\begin{equation}\label{eq:scoef_binned_complete}
 \begin{split}
  \scoef_{ij}
  &= \sqrt{\ki\kmi\kj\kmj}\left[
    \lbr\cj\ci + \sj\si\rbr\sindbeta - \lbr\sj\ci - \cj\si\rbr\cosdbeta \right] \\
  &+ \rbd\kmi\sqrt{\kj\kbmj}\left[\cpj\sin{\lbr2\pphi + \gphi\rbr} -
                                  \spj\cos{\lbr2\pphi + \gphi\rbr}\right]\\
  &+ \rbd\ki\sqrt{\kmj\kbj}\left[\cpmj\sin{\lbr2\pphi+\gphi\rbr} +
                                 \spmj\cos{\lbr2\pphi+\gphi\rbr}\right]\\
  &+ \rbd^2\sqrt{\ki\kmi\kbj\kbmj}\\ 
  & \times\left[\lbr\cbj\ci - \sbj\si\rbr\sin{\lbr2\pphi+2\gphi\rbr} - 
               \lbr\sbj\ci + \cbj\si\rbr\cos{\lbr2\pphi+2\gphi\rbr}\right],
 \end{split}
\end{equation}
where
\begin{equation}\label{eq:kjkmj}
 \kj = \int\limits_{\mcb_j}\pbdbar\bvar \mathrm{d}\mu^2_{+}\mu^2_{-},\quad
 \kbj = \int\limits_{\mcb_j}\pbd\bvar \mathrm{d}\mu^2_{+}\mu^2_{-},
\end{equation}
\begin{subequations}\label{eq:cs_complete}
 \begin{align}
  \cj  + i\sj  &= \frac{1}{\sqrt{\kj\kmj}}\int\limits_{\mcb_j}
                  \abdbar^*\bvar\abdbar\bvarinv\mathrm{d}\mu^2_{+}\mu^2_{-},\\
  \cbj + i\sbj &= \frac{1}{\sqrt{\kbj\kbmj}}\int\limits_{\mcb_j}
                  \abd^*\bvar\abd\bvarinv\mathrm{d}\mu^2_{+}\mu^2_{-}, \\
  \ctj + i\stj &= \frac{1}{\sqrt{\kj\kbj}}\int\limits_{\mcb_j}
                  \abdbar^*\bvar\abd\bvar\mathrm{d}\mu^2_{+}\mu^2_{-}, \\
  \cpj + i\spj &= \frac{1}{\sqrt{\kj\kbmj}}\int\limits_{\mcb_j}
                  \abdbar^*\bvar\abd\bvarinv\mathrm{d}\mu^2_{+}\mu^2_{-}
 \end{align}
\end{subequations}

Definitions in Eq.~(\ref{eq:cs_complete}) imply
\begin{equation}\label{eq:cs_complete_symmetries}
  \cmj + i\smj \equiv \cj - i\sj,\quad
  \cbmj + i\sbmj \equiv \cbj - i\sbj.
\end{equation}

The expressions for $\cpconj$ specific $D$ meson decays and $\btodh$ decay 
can be obtained as a particular cases of Eqs.~(\ref{eq:ucoef_complete}),~(\ref{eq:ccoef_complete}) 
and~(\ref{eq:scoef_complete}):

\begin{itemize}
 \item $\bdcppp$
 \begin{equation}\label{eq:u_bdpp_dcp_scs}
  \begin{split}
   \ucoef_{j} &= \frac{1}{2}\lbr \kj + \kmj \rbr + \frac{1}{2}\rbd^2\lbr\kbj + \kbmj\rbr\\
              &+ \rbd\xi_D\left[\sqrt{\kj\kbj}\lbr \ctj\cos{\gphi} - \stj\sin{\gphi} \rbr 
              +  \sqrt{\kmj\kbmj}\lbr \ctmj\cos{\gphi} + \stmj\sin{\gphi} \rbr \right],
  \end{split}
 \end{equation}
 \begin{equation}\label{eq:c_bdpp_dcp_scs}
  \begin{split}
   \ccoef_{j} &= \frac{1}{2}\lbr \kj - \kmj\rbr + \frac{1}{2}\rbd^2\lbr\kbj - \kbmj\rbr \\
              &+ \rbd\xi_D\left[\sqrt{\kj\kbj}\lbr \ctj\cos{\gphi} - \stj\sin{\gphi} \rbr 
              -  \sqrt{\kmj\kbmj}\lbr \ctmj\cos{\gphi} + \stmj\sin{\gphi} \rbr \right],
  \end{split}
 \end{equation}
 \begin{equation}\label{eq:s_bdpp_dcp_scs}
  \begin{split}
   \scoef_{j}
   &=\xi_D\sqrt{\kj\kmj}\lbr \cj\sindbeta - \sj\cosdbeta \rbr \\
   &+\rbd\sqrt{\kj\kbmj}\left[\cpj\sin{\lbr2\pphi + \gphi\rbr} -
                              \spj\cos{\lbr2\pphi + \gphi\rbr}\right]\\
   &+\rbd\sqrt{\kmj\kbj}\left[\cpmj\sin{\lbr2\pphi+\gphi\rbr} +
                              \spmj\cos{\lbr2\pphi+\gphi\rbr}\right]\\
   &+\rbd^2\xi_D\sqrt{\kbj\kbmj}\left[ \cbj\sin{\lbr 2\pphi + 2\gphi\rbr} -
                                       \sbj\cos{\lbr 2\pphi + 2\gphi\rbr} \right].
  \end{split}
 \end{equation}
 \item $\btodh$, $\dkpp$
 \begin{subequations}\label{eq:ud_bdh_dkpp_scs}
  \begin{align}
   \label{eq:u_bdh_dkpp_scs}
   \ucoef_{i} &= \frac{1 + \rbd^2}{2}\lbr \kmi + \ki \rbr
   + 2\rbd\cos{\dbd}\sqrt{\ki\kmi}\lbr \ci\cos{\gphi} + \si\sin{\gphi} \rbr, \\ 
   \label{eq:c_bdh_dkpp_scs}
   \ccoef_{i} &= \frac{1 - \rbd^2}{2}\lbr \kmi - \ki \rbr
   + 2\rbd\sin{\dbd}\sqrt{\ki\kmi}\lbr \si\cos{\gphi} - \ci\sin{\gphi} \rbr,
  \end{align}
 \end{subequations}
 \begin{equation}\label{eq:s_bdh_dkpp_scs}
  \begin{split}
   \xi_{h^0}\scoef_{i} 
    &= \sqrt{\ki\kmi}\lbr \ci\sindbeta + \si\cosdbeta \rbr \\
    &+\rbd\left[\ki\sin{\lbr2\pphi + \gphi + \dbd\rbr} +
                              \kmi\sin{\lbr2\pphi + \gphi - \dbd\rbr} \right] \\
    &+\rbd^2\left[\ci\sin{\lbr2\pphi + 2\gphi\rbr} -
                                 \si\cos{\lbr2\pphi + 2\gphi\rbr}\right],
  \end{split}
 \end{equation}
 where the coefficient $\xi_{h^0}\equiv\lbr -1 \rbr^L\xi_{\cpconj}^{h^0}$ accounts for 
 the $\cpconj$ parity of $h^0$ meson and the angular moment $L$ of the $Dh^0$ system.
 \item $\bdcph$
 \begin{subequations}\label{eq:ucs_bdh_dcp_scs}
 \begin{align}
  \label{eq:u_bdh_dcp_scs}
  \ucoef &= 1 + \rbd^2 + 2\xi_D\rbd\cos{\dbd}\cos{\gphi}, \\ 
  \label{eq:c_bdh_dcp_scs}
  \ccoef &= - 2\xi_D\rbd\sin{\dbd}\sin{\gphi}, \\
  \label{eq:s_bdh_dcp_scs}
  \xi_{h^0}\scoef &= \xi_D\sindbeta + 2\rbd\cos{\dbd}\sin{\lbr 2\pphi + \gphi \rbr}
 + \rbd^2\xi_D\sin{\lbr2\pphi+2\gphi\rbr}.
 \end{align}
\end{subequations}
\end{itemize}

As discussed in Ref~\cite{gronau_bdks}, the expressions~(\ref{eq:ud_bdh_dkpp_scs}), 
(\ref{eq:s_bdh_dkpp_scs}), and (\ref{eq:ucs_bdh_dcp_scs}) describe also the 
time-dependent analysis of tagged $\bn\to D\ks$ decays.  The $\ckm$ angles~$\pphi$ 
and~$\gphi$, phase~$\delb$ and parameter~$\rbd$ can be simultaneously measured in 
a such analysis.  In contrast with the~$\bdh$ decay, the~$\rbd$ value corresponding 
to the~$\bn\to D\ks$ decay can be as large as~$0.2$, improving sensitivity to the~$\cpconj$ 
violation parameters.  However, the expected number of reconstructed 
at a~$B$ factory $\bn\to D\ks$ decays is about the order of magnitude less then the 
number of reconstructed $\bdh$ decays.  Numerical experiments have been performed 
to estimate the statistical precision one may expect with the~$\belleii$ data.  The results obtained with $\rbd=0.2$ are
\begin{equation}\label{eq:precision_dks}
 \sigma^{(\bn\to D\ks)}(\pphi)\approx 5\grad,\quad \sigma^{(\bn\to D\ks)}(\gphi)\approx 8\grad.
\end{equation}
These values are only marginally dependent on~$\delb$. The angle~$\gphi$ 
precision doesn't improve much if the~$\pphi$ value is considered as known.

\section{\boldmath Formalism accounting for the charm mixing}\label{app:charm-mixing}
We assume conservation of $\cpconj$ symmetry in charm mixing.  The $\bdpp$, $\dkpp$
decay amplitude taking into account charm mixing can be written as follows:
\begin{equation}\label{eq:time-pdf-mixing-ampl}
 \begin{split}
  \mca_{B\to f}&\lbr \dt, t_D, \mu^2_{+},\mu^2_{-}, \mpsq, \mmsq \rbr \\
  &= \left[\ad\dvarinv\dkap + \ad\dvar    i\dsig\right] \\
  &\quad\quad\quad\quad\quad\quad\quad\quad\quad\quad\quad\quad
  \times \abdbar\bvar     \cos\hdmdt \\
  &+ \left[\ad\dvar   \dkap + \ad\dvarinv i\dsig\right] \\
  &\quad\quad\quad\quad\quad\quad\quad\quad\quad\quad\quad\quad
  \times \abdbar\bvarinv i\sin\hdmdt e^{-2i\pphi},
 \end{split}
\end{equation}
where $t_D$ is the $D$ meson proper decay time and functions
\begin{equation}\label{eq:kappa}
  \varkappa\lbr t_D\rbr = e^{-\frac{t_D}{2\tau_D}}
     \cos{\left[\frac{t_D\lbr x- iy\rbr}{2\tau_D}\right]}\quad
  \mathrm{and}\quad
  \sigma\lbr    t_D \rbr = e^{-\frac{t_D}{2\tau_D}}
     \sin{\left[\frac{t_D\lbr x- iy\rbr}{2\tau_D}\right]}
\end{equation}
describe the $D$ meson time evolution. Here $x$ and $y$ are the charm mixing 
parameters and $\tau_D$ is the $\dn$ lifetime.  The corresponding amplitude 
of the $\bnbar\to D\pipi$, $\dkpp$ decay is
\begin{equation}\label{eq:time-pdf-mixing-ampl-bar}
 \begin{split}
  \mca_{\bbar\to f}&\lbr \dt, t_D, \mu^2_{+},\mu^2_{-}, \mpsq, \mmsq \rbr \\
  &= \left[\ad\dvar   \dkap + \ad\dvarinv i\dsig\right] \\
  &\quad\quad\quad\quad\quad\quad\quad\quad\quad\quad\quad\quad
  \times \abdbar\bvarinv  \cos\hdmdt \\
  &+ \left[\ad\dvarinv\dkap + \ad\dvar    i\dsig\right] \\
  &\quad\quad\quad\quad\quad\quad\quad\quad\quad\quad\quad\quad
  \times \abdbar\bvar    i\sin\hdmdt e^{2i\pphi},
 \end{split}
\end{equation}

The coefficients $\ucoef$, $\ccoef$, and $\scoef$, defined in 
Eq.~(\ref{eq:time-pdf-udf}) corresponding to amplitudes in 
Eqs.~(\ref{eq:time-pdf-mixing-ampl}) and (\ref{eq:time-pdf-mixing-ampl-bar}), 
integrated over the $D$ meson proper decay time $t_D$, are
\begin{equation}\label{eq:u_mixing_unbinned}
 \begin{split}
  \ucoef &=
    \frac{1}{4}\lbr \frac{1}{1-y^2} + \frac{1}{1 + x^2} \rbr
       \left[p_B\bvar p_D\dvarinv + p_B\bvarinv p_D\dvar \right] \\
  &+\frac{1}{4}\lbr \frac{1}{1-y^2} - \frac{1}{1 + x^2} \rbr
       \left[p_B\bvar p_D\dvar    + p_B\bvarinv p_D\dvarinv \right] \\
  &+\sqrt{p_D\dvar p_D\dvarinv} \\ &\times
        \left(\frac{1}{2}\frac{x}{1 + x^2}\,\sin{\deld}
              \left[p_B\bvar - p_B\bvarinv\right] \right. \\
  &+\left.\ \ \frac{1}{2}\frac{y}{1 - y^2}\,\cos{\deld}
              \left[p_B\bvar + p_B\bvarinv\right]\right),
 \end{split}
\end{equation}

\begin{equation}\label{eq:c_mixing_unbinned}
 \begin{split}
  \ccoef &=
    \frac{1}{4}\lbr \frac{1}{1-y^2} + \frac{1}{1 + x^2} \rbr
       \left[p_B\bvar p_D\dvarinv - p_B\bvarinv p_D\dvar \right] \\
  &+\frac{1}{4}\lbr \frac{1}{1-y^2} - \frac{1}{1 + x^2} \rbr
       \left[p_B\bvar p_D\dvar    - p_B\bvarinv p_D\dvarinv \right] \\
  &+\sqrt{p_D\dvar p_D\dvarinv} \\ &\times
        \left(\frac{1}{2}\frac{x}{1 + x^2}\,\sin{\deld}
        \left[p_B\bvar + p_B\bvarinv\right] \right. \\
  &+\left.\ \ \frac{1}{2}\frac{y}{1 - y^2}\,\cos{\deld}
        \left[p_B\bvar - p_B\bvarinv\right]\right),
 \end{split}
\end{equation}

\begin{equation}\label{eq:s_mixing_unbinned}
 \begin{split}
  \scoef &=
  \sqrt{p_D\dvar p_D\dvarinv p_B\bvar p_B\bvarinv} \\
  &\times \left[\frac{1}{2}\lbr \frac{1}{1-y^2} + 
      \frac{1}{1 + x^2} \rbr\sin{\lbr2\pphi - \deld + \dbd\rbr}\right. \\
  &+\,\left.\ \frac{1}{2}\lbr \frac{1}{1-y^2} -
      \frac{1}{1 + x^2} \rbr\sin{\lbr2\pphi - \deld - \dbd\rbr}\right] \\
  &-\sqrt{p_B\bvar p_B\bvarinv} \\ &\times
        \left(\frac{1}{2}\frac{y}{1 - y^2}\sin{\lbr2\pphi - \dbd\rbr}
        \left[p_D\dvar + p_D\dvarinv\right] \right. \\
  &-\left.\ \ \frac{1}{2}\frac{x}{1 + x^2}\cos{\lbr2\pphi - \dbd\rbr}
        \left[p_D\dvar - p_D\dvarinv\right]\right).
 \end{split}
\end{equation}

Integrating Eqs.~(\ref{eq:u_mixing_unbinned}),~(\ref{eq:c_mixing_unbinned}) 
and~(\ref{eq:s_mixing_unbinned}) over $i^{\mathrm{th}}$ bin of the $D$ Dalitz 
plot and $j^{\mathrm{th}}$ bin of the $B$ Dalitz plot we obtain the expressions for
the binned analysis:
\begin{subequations}\label{eq:uc_mixing}
 \begin{align}
  \label{eq:u_mixing}
  \ucoef_{ij} &= \frac{1}{2}\kj\kmi + \frac{1}{2}\kmj\ki
              + \frac{1}{2}\sqrt{\ki\kmi}
              \left[y\ci\lbr \kj + \kmj \rbr + x\si\lbr \kj - \kmj \rbr\right], \\ 
  \label{eq:c_mixing}
  \ccoef_{ij} &= \frac{1}{2}\kj\kmi - \frac{1}{2}\kmj\ki
              + \frac{1}{2}\sqrt{\ki\kmi}
              \left[y\ci\lbr \kj - \kmj \rbr + x\si\lbr \kj + \kmj \rbr\right],
 \end{align}
\end{subequations}
\begin{equation}\label{eq:s_mixing}
 \begin{split}
  \scoef_{ij} &= \sqrt{\kj\kmj\ki\kmi}\left[\lbr \ci\cj+\si\sj \rbr\sindbeta - 
                                            \lbr \ci\sj-\si\cj \rbr\cosdbeta\right] \\
              &+ \frac{1}{2}\sqrt{\kj\kmj}\left[
               y\lbr \sj\cosdbeta - 
                     \cj\sindbeta \rbr
                     \lbr \ki + \kmi \rbr \phantom{\frac{1}{2}} \right.\\
              &\left.\phantom{\frac{1}{2}\sqrt{\kj\kmj}}\ 
              + x\lbr \cj\cosdbeta + 
                      \sj\sindbeta \rbr \lbr \ki - \kmi \rbr
              \right].
 \end{split}
\end{equation}

The expressions for $\cpconj$ specific $D$ meson decays and $\btodh$ decay 
can be obtained as a particular cases of Eqs.~(\ref{eq:uc_mixing}) and~(\ref{eq:s_mixing}):
\begin{itemize}
 \item $\bdcppp$
 \begin{subequations}\label{eq:ucs_bdpp_dcp_mixing}
 \begin{align}
  \label{eq:u_bdpp_dcp_mixing}
  \ucoef_{j} &= \frac{1}{2}\lbr\kj + \kmj\rbr\lbr 1+ \xi_D y\rbr, \\ 
  \label{eq:c_bdpp_dcp_mixing}
  \ccoef_{j} &= \frac{1}{2}\lbr\kj - \kmj\rbr\lbr 1+ \xi_D y\rbr, \\
  \label{eq:s_bdpp_dcp_mixing}
  \scoef_{j} &= \xi_D\sqrt{\kj\kmj}\lbr\cj\sindbeta - \sj\cosdbeta\rbr\lbr 1 - \xi_D y\rbr.
 \end{align}
\end{subequations}
 \item $\btodh$, $\dkpp$
 \begin{subequations}\label{eq:uc_bdh_mixing}
 \begin{align}
  \label{eq:u_bdh_mixing}
  \ucoef_{i} &= \frac{1}{2}\lbr \kmi + \ki \rbr + y\ci\sqrt{\ki\kmi}, \\ 
  \label{eq:c_bdh_mixing}
  \ccoef_{i} &= \frac{1}{2}\lbr \kmi - \ki \rbr + x\si\sqrt{\ki\kmi},
 \end{align}
\end{subequations}
\begin{equation}\label{eq:s_bdh_mixing}
 \begin{split}
  \xi_{h^0}\scoef_{i} &= \sqrt{\ki\kmi}\lbr\ci\sindbeta - \si\cosdbeta\rbr \\
             &+ \frac{1}{2}\left[x\cosdbeta \lbr \ki - \kmi \rbr -
                                 y\sindbeta \lbr \ki + \kmi \rbr \right].
 \end{split}
\end{equation}
 \item $\bdcph$
 \begin{subequations}\label{eq:ucs_bdh_dcp_mixing}
 \begin{align}
  \label{eq:u_bdh_dcp_mixing}
  \ucoef &= 1 + \xi_D y, \\ 
  \label{eq:c_bdh_dcp_mixing}
  \ccoef &= 0, \\
  \label{eq:s_bdh_dcp_mixing}
  \xi_{h^0}\scoef &= \xi_D\sindbeta\lbr 1 - \xi_D y \rbr.
 \end{align}
\end{subequations}
\end{itemize}

\section{\boldmath Estimate of the bias due to neglect of 
\texorpdfstring{$\btoucd$}{b -> c ubar d} transition and charm mixing}
\label{app:systematics-evaluation}
The neglect of $\btoucd$ transition and charm mixing leads to a
bias of the observed value of the angle $\pphi$.  Numerical experiments 
have been performed to assess the bias value.  Data samples for the numerical 
experiments are generated using the expressions from appendices~\ref{app:bucbard} 
and~\ref{app:charm-mixing}.  The values of angle $\pphi$ and hadronic 
parameters~$\cj$ and~$\sj$ are extracted from the generated 
samples with the maximum likelihood method.  The fit procedure uses equations from 
Sec.~\ref{sec:bins} (i.e. neglects the $\btoucd$ transition and charm 
mixing). The results obtained are summarized in the Table~\ref{tab:beta_bias}.

\begin{table}[t]
 \caption{Estimates for the angle $\pphi$ measurement bias due
 to the neglect of $\btoucd$ transition ($3^{\mathrm{rd}}$ column) and charm mixing 
 ($4^{\mathrm{th}}$ column).  The second column shows the $\dn$ decays combination 
 used in the fit: $n_D, n_+, n_-$ are relative fractions of 
 $\dnkpp$, $\dn\to f_{\cpconj+}$ and $\dn\to f_{\cpconj-}$ decays yields, respectively.}
  \label{tab:beta_bias}
  \centering
 \begin{tabular}{lccc}
  \hline\hline
   \multicolumn{1}{c}{Process} &
   \multicolumn{1}{c}{$\lbr n_D, n_+, n_-\rbr$} &
   \multicolumn{1}{c}{$\delta\pphi_{\btoucd}$} &
   \multicolumn{1}{c}{$\delta\pphi_{\mathrm{mix}}$}
   \\ \hline
   \multirow{4}{*}{$\bdpp$} 
   & $\lbr 2, 1, 1 \rbr$ & $0.17\grad\times\frac{\rbd}{0.02}$ & $0.05\grad\times \frac{\sqrt{x^2+y^2}}{0.01}$ \\
   & $\lbr 1, 0, 0 \rbr$ & $0.15\grad\times\frac{\rbd}{0.02}$ & $0.04\grad\times \frac{\sqrt{x^2+y^2}}{0.01}$ \\
   & $\lbr 1, 1, 0 \rbr$ & $0.14\grad\times\frac{\rbd}{0.02}$ & $0.04\grad\times \frac{\sqrt{x^2+y^2}}{0.01}$ \\
   & $\lbr 1, 0, 1 \rbr$ & $0.23\grad\times\frac{\rbd}{0.02}$ & $0.05\grad\times \frac{\sqrt{x^2+y^2}}{0.01}$ \\
   \hline
   \multirow{3}{*}{$\bdh$}   
   & $\lbr 1, 0, 0 \rbr$ & $-0.2\grad\times\cos{\delb}$ & $0.02\grad\times \frac{\sqrt{x^2+y^2}}{0.01}$ \\
   & $\lbr 0, 1, 0 \rbr$ & $-1.9\grad\times\cos{\delb}$ & $-0.6\grad\times \frac{y}{0.01}$ \\
   & $\lbr 0, 0, 1 \rbr$ & $\phantom{-}1.9\grad\times\cos{\delb}$ & $\phantom{-}0.6\grad\times \frac{y}{0.01}$ \\
  \hline\hline
 \end{tabular}
\end{table}

A model of the suppressed $\bdnpp$ 
decay is needed to obtain the values of parameters 
$\kbj$, $\cbj$, $\sbj$, $\ctj$, $\stj$, $\cpj$ and~$\spj$ 
defined in Eqs.~(\ref{eq:kjkmj}) and~(\ref{eq:cs_complete}).  We use  the  
factorization assumption\footnote{The factorization assumption 
is not applicable to the $\bdnpp$ decay, but it gives a qualitative arguments 
to construct the $\bdnpp$ decay model as described in text.} 
to construct an ensemble of the $\bdnpp$ decay models. The $\bdbpp$ decay model 
described in appendix~\ref{app:decay_ampl} is taken as a basis and the following 
modifications are applied:
\begin{itemize}
 \item The $\bn\to D_2^{*-}(2460)\pi^+$ transition amplitude is reduced by a 
 factor of~$10$ since it cannot proceed through a tree weak diagram.
 \item The $B^0\to R\pi^-$, $R\to\dnbar\pi^+$, $R\in\{D^*_2, D^v_*, D_0^*\}$ 
 amplitudes are increased by factor $f_D / f_{\pi} \approx 1.6$, where 
 $f_D\approx207\ \mev$ and $f_{\pi}\approx133\ \mev$ are the decay constants.
 \item The amplitudes of $\bn\to \dnbar R$, $R\to\pipi$ transitions are taken from the $\bdbpp$ 
 decay model since the production mechanisms of the $\pi^+\pi^-$ resonances in $\bdnpp$ and
 $\bdbpp$ decays are similar.
\end{itemize}
An ensemble of $100$ $\bdnpp$ decay models is constructed with 
$100$ random triples of phases $\varphi(D^*_2), \varphi(D^v_*), \varphi(D_0^*)$ 
corresponding to the $D^*_2$, $D^v_*$ and $D_0^*$ amplitudes, respectively.  
The values quoted in the third column of Table~\ref{tab:beta_bias} for the 
$\bdbpp$ decay are the maximal biases over the ensemble of models.

The results of numerical experiments and formalism described in 
appendices~\ref{app:bucbard} and~\ref{app:charm-mixing} lead to 
the following conclusions:
\begin{enumerate}
 \item The bias due to neglect of the charm mixing is $3\div 4$ times 
 smaller than the bias due to neglect of the $\btoucd$ transition.
 
 \item The biases corresponding $\cpconj$ specific $D$ meson final states 
 with~$\xi_D=+1$ and~$\xi_D=-1$ have equal absolute values and opposite 
 signs.  This feature was previously pointed  out in Ref.~\cite{fleischer}.  
 Eqs.~(\ref{eq:s_bdpp_dcp_scs}),~(\ref{eq:s_bdh_dcp_scs}),~(\ref{eq:s_bdpp_dcp_mixing}) 
 and~(\ref{eq:s_bdh_dcp_mixing}) show that the main terms are proportional 
 to the $\xi_D$ while the first order corrections do not depend on $\xi_D$.
 
 \item Biases for the processes involving $\dkpp$ decay are of the order of 
 $0.1\grad$.  Relative smallness of this value can be qualitatively 
 explained by the pairwise reduction of bias in  bins of the Dalitz 
 plot.  This effect generalizes the feature described in the previous item.  
 The same reduction is takes place in the binned analysis of $\bdbpp$ decay.
 
 \item The biases for the $\bn\to D_{\cpconj}h^0$ decays are large enough 
 to be observed with the $\belleii$ statistics.  However, assuming the 
 statistics ratio $2/1/1$ of the $\kspp$, $\xi_D=+1$ and $\xi_D=-1$ 
 events, respectively (which is close to reality), the residual bias is 
 about $0.1\grad$.
 
 \item Most of the $\dn$ decays to $\cpconj$ eigenstates collected by 
 $\lhcb$ have negative $\cpconj$ parity ($\dn\to K^+K^-, \pi^+\pi^-$).  
 This $\cpconj$ parity imbalance does not lead to a significant 
 bias in the case of analysis of the $\bdbpp$ decays, in contrast with 
 the $\bdh$ case, as it is shown in the third and fourth rows of the 
 Table.~\ref{tab:beta_bias}.  The resulting bias due to neglect of the $\btoucd$ 
 amplitude is at level of~$0.2\grad$.
\end{enumerate}

\end{document}